\documentclass[a4paper]{article}

\usepackage{amsmath,amssymb}
\usepackage
{graphicx}
\usepackage{color}
\usepackage{cite}

\usepackage{setspace}

\newcommand{\vect}[1]
{{\mbox{\boldmath $#1$}}}

\newcommand{\svect}[1]
{{\mbox{\boldmath{\scriptsize $#1$}}}}

\makeatletter

\@addtoreset{equation}{section}
\makeatother

\newcommand{\beq}
{\begin{eqnarray}}

\newcommand{\eeq}
{\end{eqnarray}}

\newcommand{\Gam}
{{\it\Gamma}}

\newcommand{\RR}
{\mathrm{Re}\,}

\newcommand{\II}
{\mathrm{Im}\,}

\begin{document}

\quad
\\

\begin{center}
{\bf\LARGE{QFT in the flat chart of de Sitter space}}\\
\end{center}
\quad 
\quad \\

\begin{center}
{\large{Yusuke Korai and Takahiro Tanaka\\
\quad \\
{\it{Yukawa Institute for Theoretical Physics, Kyoto University,\\
Kyoto, 606-8502, Japan}}\\

\quad\\
\href{mailto:korai@yukawa.kyoto-u.ac.jp}
{\texttt{korai@yukawa.kyoto-u.ac.jp}},\quad
\href{mailto:tanaka@yukawa.kyoto-u.ac.jp}
{\texttt{tanaka@yukawa.kyoto-u.ac.jp}}
}
}
\end{center}

\quad\\
\quad \\

\begin{abstract}

We study the correlators for interacting quantum field theory in
the flat chart of de Sitter space at all orders in perturbation.
The correlators are calculated in the in-in formalism which are often 
applied to the calculations in the cosmological perturbation.
It is shown that these correlators are de Sitter invariant.
They are compared with the correlators calculated based 
on the Euclidean field theory. We then find that these two 
correlators are identical.
This correspondence has been already shown
graph by graph
but we give an alternative proof of it by direct calculation.

\end{abstract}
\newpage
\tableofcontents

\section{Introduction}

In recent years, there have been rapid progresses 
in the precise measurement of the observable quantities in cosmology, 
e.g., the non-Gaussianity of the fluctuations generated during inflation,
which is expected to be a powerful tool as a probe of the early universe.
Along with the development of these precise measurements, 
the need arises for the accurate theoretical predictions 
of the corresponding quantities. 

When computing the non-Gaussianity, one needs to discuss 
interacting quantum field theory on inflationary background, 
in which one does not generally know how to define the interacting vacuum.
One often uses the $i\epsilon$ prescription 
in cosmology to calculate the correlators 
perturbatively. (See, for example, Ref.~\cite{Maldacena:2002vr}.) 
In the Minkowski, 
this prescription is known to perturbatively give 
the Poincar\'{e} invariant correlators for interacting theory, 
defining interacting vacuum as the lowest energy eigen state. 
Indeed, this prescription also enables us to calculate 
the non-Gaussianity or higher correlations in the inflationary era, 
but the physical meaning of it is not as clear as in the Minkowski case. 
Our main interest in this paper is in the meaning of 
the $i\epsilon$ prescription for interacting field theory in de Sitter space.

The free scalar quantum field theory in de Sitter space 
is well understood 
\cite{Allen:1985ux, Allen:1987tz, Folacci:1992xc, Kirsten:1993ug}, 
while interacting one is a hot subject with a lot of debate 
\cite{Ford:1977dj, Ford:1984hs, 
Higuchi:2008tn,
Marolf:2010zp, Marolf:2010nz, Higuchi:2010xt, Marolf:2011sh, Higuchi:2011vw,
Marolf:2012kh, 
Hollands:2010pr,Hollands:2011we,
Rajaraman:2010xd,
Polyakov:2007mm,Polyakov:2009nq,Krotov:2010ma,
Tsamis:1994ca,Tsamis:1996qq,
Garriga:2007zk, 
Boyanovsky:2004gq,
Onemli:2004mb, Kahya:2006hc, Kahya:2010xh, 
Giddings:2011zd, Giddings:2011ze,
Kitamoto:2011yx, Kitamoto:2012ep, Kitamoto:2012vj,
Urakawa:2009my, Urakawa:2010it, Urakawa:2010kr, Tanaka:2011aj,
Einhorn:2002nu,
Weinberg:2005vy, Weinberg:2006ac, Weinberg:2010wq, 
Senatore:2009cf, Senatore:2012nq, Pimentel:2012tw}. 
We focus in the present paper 
on the problem whether the $i\epsilon$ prescription
for interacting theory breaks de Sitter invariance.

Since de Sitter space is maximally symmetric 
and possesses $SO(4,1)$, de Sitter symmetry, it is strongly expected 
to exist a de Sitter invariant vacuum even for interacting theory. 
In fact, a de Sitter invariant vacuum for interacting theory is defined 
by constructing arbitrary correlators perturbatively 
at all orders using the Euclidean method \cite{Marolf:2010nz}.
While the vacuum state thus constructed 
are manifestly de Sitter invariant,
it is not obvious whether the ones defined 
by the $i\epsilon$ prescription in the flat chart 
are de Sitter invariant. 
Notice, for example, that in the latter the integration region 
for the vertices in calculating correlators 
are restricted to the future of the cosmological horizon, 
which is not de Sitter invariant.

Actually, this problem has been already resolved affirmatively 
in Ref.~\cite{Higuchi:2010xt} for interacting massive scalar field. 
Namely, the $i\epsilon$ prescription does not break de Sitter invariance
for interacting massive scalar field.
Furthermore, the vacuum defined by the $i\epsilon$ prescription
has been shown to be equivalent to the Euclidean vacuum.
The main ideas in Ref.~\cite{Higuchi:2010xt} are as follows. They start from correlators 
defined on an Euclidean sphere and take, on the Euclidean sphere, 
coordinates such that when we Wick rotate the time coordinate 
continues to the static chart of the Lorentzian de Sitter space. 
Then, after the deformation of the integral path of the Euclidean time, 
fall-off of the propagator in the large separation limit leads 
to the identity of the two correlators at least on the static chart. 
From the analyticity of the in-in correlators for their time coordinates, 
and the uniqueness of the analytic continuation, it is shown that 
the in-in correlators in the flat chart are identical to the analytic 
continuations of those on an Euclidean shpere.

Then, it is natural to ask what becomes of in massless field theory.
What happens for graviton in de Sitter space has especially been 
a topic of much discussion. (See, e.g. \cite{Tsamis:1994ca, Tsamis:1996qq, 
Garriga:2007zk, Higuchi:2011vw}.) 
Our final goal is to extend the correspondence between the two vacua 
to those interacting massless field theory. It is also worth considering
derivatively interacting massless scalar field,
which can be a step toward graviton.

It seems difficult to extend the discussion of massive field theory above
to massless field theory where the propagator does not fall off in general,
since the proof of the correspondence between 
the two vacua relies on this decay property
of the propagator at a large separation as explained above.
In order to attack those theories, we take another approach.
That is, we directly calculate the correlators with the $i\epsilon$ prescription.
We derive, along this way, the analytic Mellin-Barnes 
formulae for the correlators of quantum fields in the flat chart. 
The resulting correlators are shown to be 
completely the same as the analytic continuations of 
the ones considered in the Euclidean field theory in Ref.~\cite{Marolf:2010nz}. 
Thus we find that the $i\epsilon$ prescription 
in de Sitter space gives the vacuum state 
corresponding to the Euclidean field theory.
Although we consider only massive theory in the present paper,
we believe that our proof has potential to be extended to
wider range of theories which include interacting massless theory 
such as derivatively interacting one,
since it does not employ the decay property 
of the propagator.

This paper is organized as follows. In Sec.~\ref{sec2}, 
we briefly review how to describe de Sitter space, 
especially the flat chart, and massive free scalar quantum field theory on it. 
Pauli-Villars regularization scheme is also introduced. 
Then we proceed to the interacting theory, 
in Sec.~\ref{sec3}, \ref{sec4} and \ref{sec5}. 
We consider, in Sec.~\ref{sec3} and \ref{sec4}, a tree graph which contributes to an 
$N$-pt correlator with single vertex. Then in Sec.~\ref{sec5}, we extend the 
discussion to arbitrary graphs. 
We give a brief summary in Sec.~\ref{sec6}.

\section{\label{sec2}Preliminaries}

In this section, we briefly review free scalar quantum field theory on 
de Sitter space, especially in the flat chart. 
We also introduce Pauli-Villars regularization scheme for later use. 

\subsection{de Sitter space}

We consider $D$-dimensional de Sitter space $dS^{D}$ with, for
simplicity, unit radius. This is a hyperboloid embedded in
$(D+1)$-dimensional Minkowski space with metric $\eta_{ab}=(-,+,\cdots,+)$. 
The embedding is specified by 
\beq
\eta_{ab}X^{a}X^{b}=1~. 
\eeq
It is convenient to define the invariant distance between two points $X$
and $Y$ in de Sitter space by the Minkowski inner product
of $X$ and $Y$, which we denote as 
\beq
Z(X,Y):=\eta_{ab}X^{a}Y^{b}~,
\eeq
as in Ref.~\cite{Marolf:2010zp}. 
For brevity, we often use alternative notation $Z_{XY}$ for 
$Z(X,Y)$, $Z_{1Y}$ for $Z(X_{1},Y)$ 
and so forth in the following. 

The coordinates $(\eta,\vect{x})$ in the flat chart are 
related to the embedding coordinates as
\beq
X^{0}=\frac{1}{2}\left(\eta-\frac{1}{\eta}\right)-\frac{||\vect{x}||^{2}}{2\eta}~,\quad
X^{D}=-\frac{1}{2}\left(\eta+\frac{1}{\eta}\right)+\frac{||\vect{x}||^{2}}{2\eta}~,\quad
X^{\alpha}=-\frac{x^{\alpha}}{\eta}~,\quad (\alpha=1,2,\cdots,D-1)~,\nonumber\\
\eeq
where $||\vect{x}||$ means the norm of $(D-1)$-vector $\vect{x}$. 
The flat chart coordinates with 
 $-\infty<\eta<0$ and $\vect{x}\in\mathbf{R}^{D-1}$ 
span just a half of the whole spacetime region. 
In fact, the linear combination 
\beq
X^{0}+X^{D}=-\frac{1}{\eta}~,
\eeq
is restricted to the positive side for negative $\eta$. 
The metric in the flat chart is expressed as
\beq
ds^{2}=\frac{1}{\eta^{2}}(-d\eta^{2}+d\vect{x}^{2})~. 
\eeq
Expressed in the flat chart coordinates, 
the invariant distance between $X$ and $X'$, $Z(X,X')$ is given by
\beq
Z(X,X')=1+\frac{(\eta-\eta')^{2}-||\vect{x}-\vect{x}'||^{2}}{2\eta\eta'}~,
\eeq
where $(\eta,\vect{x})$ and $(\eta',\vect{x}')$ are 
the flat chart coordinates corresponding to $X$ and $X'$, respectively.

\subsection{Free QFT on de Sitter}

We now consider a massive free scalar QFT on de Sitter space. 
We focus on the Green's function $G(X,Y)$ given by
\beq
G(X,Y)=\frac{\Gam(-\sigma)\Gam(\sigma+D-1)}{(4\pi)^{D/2}\Gam(D/2)}{_{2}F_{1}}
\left(-\sigma,\, \sigma+D-1;\, \frac{D}{2};\, \frac{1+Z_{XY}}{2}\right)~,
\eeq
which corresponds to taking Bunch-Davies vacuum \cite{Bunch:1978yq} or Euclidean
vacuum \cite{Gibbons:1977mu}. $\sigma$ is related to the mass of the field $m$ 
by 
\beq
\sigma=-\frac{D-1}{2}+\sqrt{\left(\frac{D-1}{2}\right)^{2}-m^{2}}~.
\eeq

Expressing the hypergeometric function in the Barnes representation, we have
\beq
\label{Barnes Green's function}
G(X,Y)=\int_{\nu}\left(\frac{1-Z_{XY}}{2}\right)^{\nu}\Gam(-\nu)\psi(\nu)~,
\eeq
with
\beq
\psi(\nu):=\frac{1}{(4\pi)^{D/2}}
\Gam\left[
\begin{matrix}
-\sigma+\nu,\, \sigma+D-1+\nu,\, 1-\frac{D}{2}-\nu\\
\frac{D}{2}+\sigma,\, 1-\frac{D}{2}-\sigma
\end{matrix}
\right]~.
\eeq
Here 
$$\Gam\left[
\begin{matrix}
\alpha_1,\, \alpha_2\,,\cdots\\
\beta_1,\, \beta_2\,,\cdots
\end{matrix}
\right]~, 
$$
stands for $\Gam(\alpha_1) \Gam(\alpha_2) \cdots/\Gam(\beta_1)
\Gam(\beta_2)\cdots$, and 
the symbol $\int_{\nu}(\cdots)$ means the Barnes integral. 
The Barnes integral is an integral along
a straight line, $C$, that traverses from $-i\infty$ to $+i\infty$ 
parallel to the imaginary axis with the factor $1/2\pi i$:
\beq
\int_{\nu}(\cdots):=\int_{C}\frac{d\nu}{2\pi i}(\cdots)~.
\eeq
The integrand of the Barnes integral 
includes sequences of poles. 
For example, $\Gam(z)$ possesses a sequence of poles
at $z=0,-1,-2,\cdots$. The integration path $C$ is taken to avoid all
the sequences of poles in the integrand. In the case
of the above Green's function, $C$ is taken to satisfy
\beq
\max\left\{-\RR\sigma-D+1,\, 
\RR\sigma\right\}<\RR\nu<\min\left\{1-\frac{D}{2},\,0\right\}~.
\eeq
This region of the integration path 
is called ``fundamental strip,'' and the poles such that are 
associated with Gamma functions like $\Gam(\cdots-\nu)$ ($\Gam(\cdots+\nu)$) and hence such
that line up on the right (left) hand side
of this strip are called right (left) poles. 
(See Fig.~\ref{poleforpsi}.) 
The symbol like
$\int_{\nu}$ is used to represent the Barnes integral in this meaning in
the following.

\subsection{Pauli-Villars Regularization}

\begin{figure}[t]
\begin{center}
\includegraphics*[height=5cm]{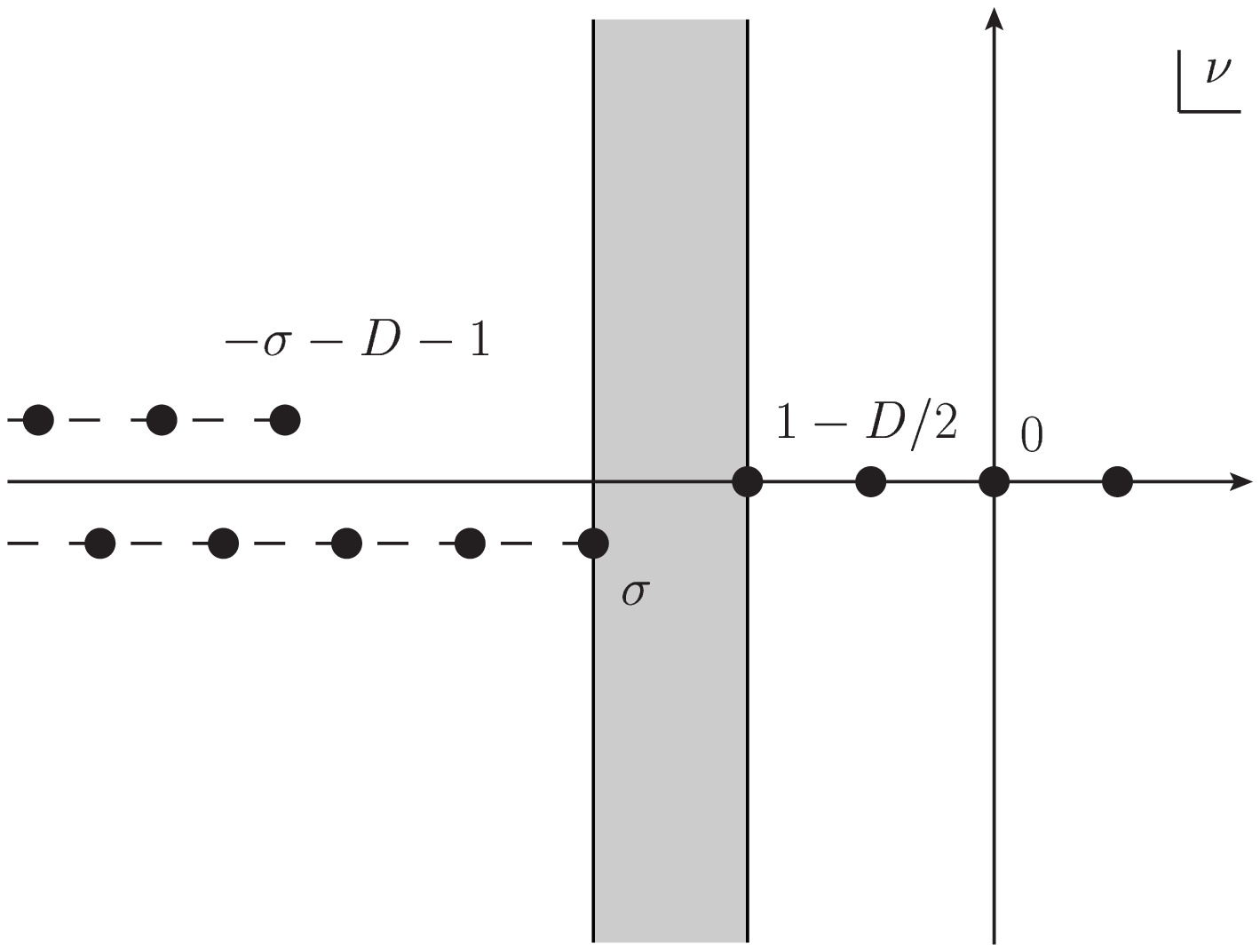}\quad \quad
\includegraphics*[height=5cm]{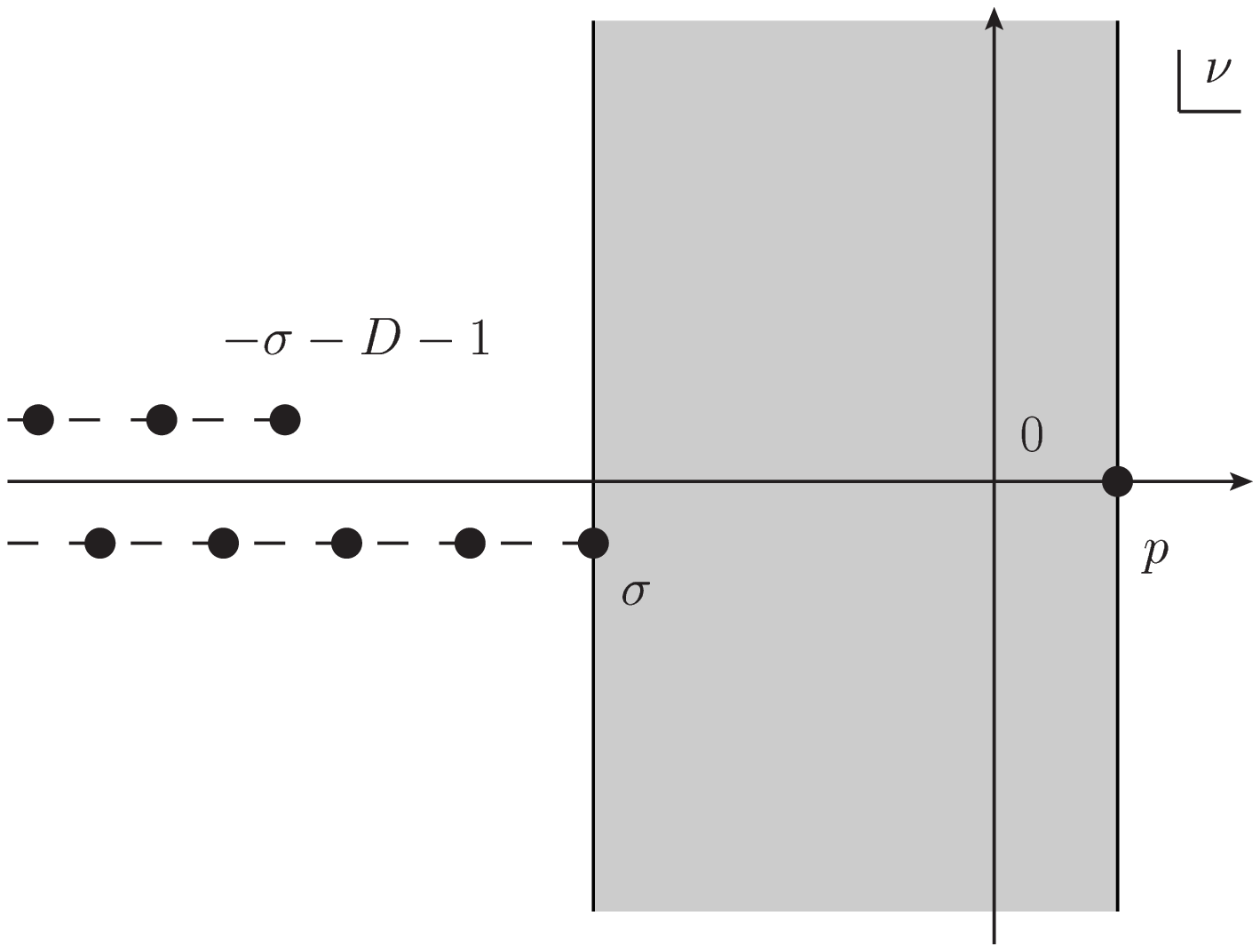}
\caption{\label{poleforpsi}
The left figure shows the pole structure for $\psi(\nu)$ which is not regularized. There are two series of left poles from $\nu=\sigma$ and $\nu=-\sigma-D-1$, and right poles from $\nu=1-D/2$. The right one shows the pole structure for $\psi^{\mathrm{reg}}(\nu)$ which is Pauli-Villars regularized. The shaded region represents the fundamental strip in each figure.}
\end{center}
\end{figure}

Because we consider interacting theory in the present paper, we have
to introduce some ultra-violet regularization scheme. We make
use of the Pauli-Villars regularization. 
This scheme attaches some massive propagators, $G_{i}(X,Y)$, 
defined in Eq.~\eqref{Barnes Green's function} with $m$ replaced 
by the regulator mass $M_{i}$, 
to the original one, $G(X,Y)$, so that we replace the original 
propagator in a graph with the regularized propagator
\beq
G^{\mathrm{reg}}(X,Y):=G(X,Y)+\sum_{i}C_{i}G_{i}(X,Y)~.
\eeq
The coefficients $C_{i}$ are chosen so that 
the regularized propagator $G^{\mathrm{reg}}(X,Y)$ becomes finite 
in the coincidence limit $Y\to X$, which leads to the conditions
\beq
\sum_{i}C_{i}=-1~,\quad \sum_{i}C_{i}M_{i}^{2}=0~,\quad \sum_{i}C_{i}M_{i}^{3}=0~,~\cdots~.
\eeq
This regularization scheme affects the pole structure of
$\psi(\nu)$ in \eqref{Barnes Green's function}, eliminating the first 
several right poles of $\psi(\nu)$ which are responsible for the 
behaviour of the Green's function in the coincidence limit \cite{Marolf:2010nz}.
The regularized Green's function is written as
\beq
G^{\mathrm{reg}}(X,Y)=\int_{\nu}\left(\frac{1-Z_{XY}}{2}\right)^{\nu}\Gam(-\nu)\psi^{\mathrm{reg}}(\nu)~,
\eeq
where we assume that $\psi^{\mathrm{reg}}(\nu)$ is regularized to
be analytic in the region 
\beq\label{analyticity region for psi}
\RR \sigma <\RR\nu< p~,
\eeq
with $p$ a sufficiently large positive constant. (See Fig.~\ref{poleforpsi}.)
In the following sections, we drop, for simplicity, the symbols such as 
$\mathrm{reg}$ on $G$ and $\psi$.

\section{\label{sec3}Interacting QFT: Single Vertex}

We now move on to the interacting theory. The interacting QFT in the flat
chart of the Lorentzian de Sitter space is discussed in the present and the
succeeding sections. When we express the correlators in the wave
number representation, we employ the $i\epsilon$ prescription to
calculate the correlators for the interacting vacuum. This prescription regularizes 
the oscillatory behaviour of the Green's functions at infinity in time
and makes the vertex integral converge. Although what we discuss in the present
paper is the position space representation of the
correlators, we also employ the $i\epsilon$ prescription to specify 
the interacting vacuum.

In this section, we discuss perturbative calculations of a single
vertex tree graph for the correlators. Then, we identify the problems 
to be solved to accomplish this calculation, which are solved
in Sec.~\ref{sec4}. In Sec.~\ref{sec5} the results for single vertex tree graphs
are extended to arbitrary graphs.

\subsection{\label{sec3.1}Definition of the In-in Path}

Let us consider $N$-pt Green's function. The contribution to $N$-pt
correlator at the lowest order in perturbation theory is given by
\beq
\label{single vertex}
\mathcal{V}_{N}(X_{1},\cdots,X_{N})
=\int_{\Omega}dV_{Y}\,G(X_{1},Y)\cdots G(X_{N},Y)~.
\eeq
In the in-in formalism with the $i\epsilon$ prescription, 
the integration region $\Omega$ for the vertex integral 
is specified as follows.

\begin{figure}[t]
\begin{center}
\includegraphics*[height=5cm]{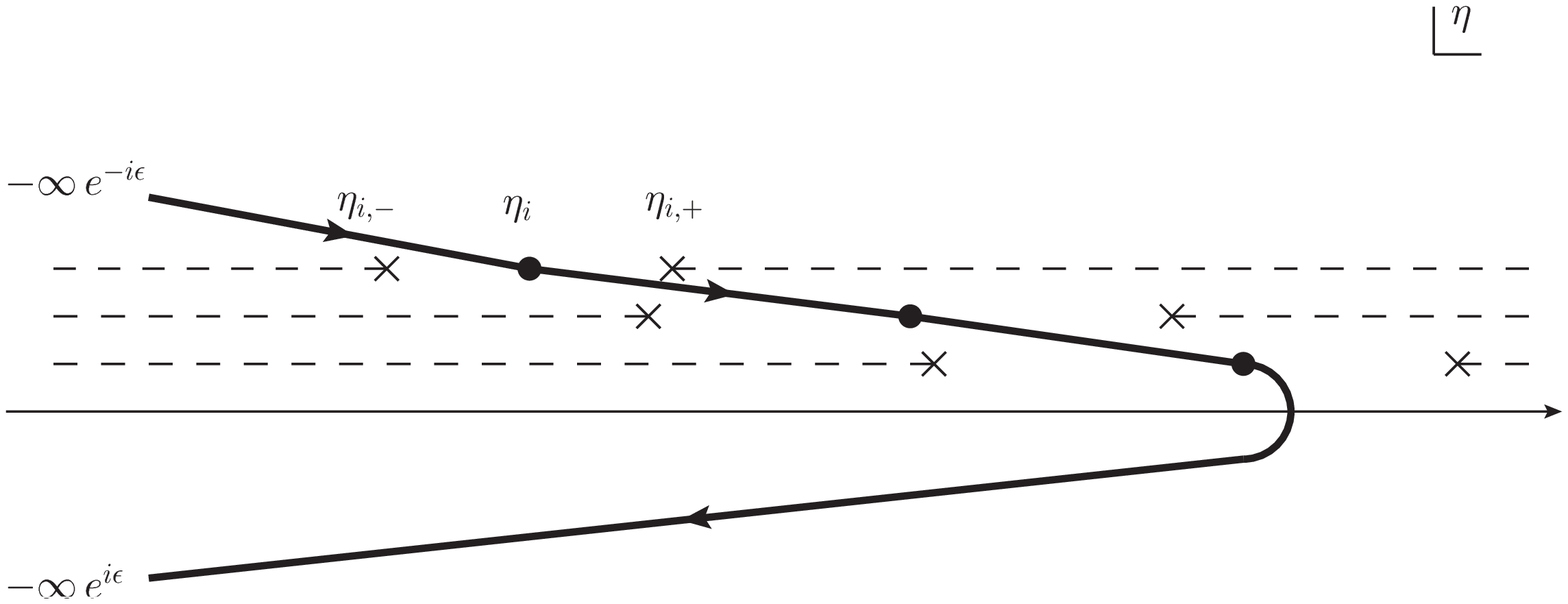}
\caption{\label{in-inpath}
This figure shows the $\eta$-path $P$ which is later deformed to $P_{\svect{y}}$. 
The dots represent the time coordinate $\eta_{i}$ of the external points 
and the crosses are the branching points corresponding to 
the light cones emanating from the external points. The dashed lines are the branch cuts.}
\end{center}
\end{figure}

We first introduce an $\eta$-integration path $P$ on the $\eta$-plane, 
independently of the spatial coordinates $\vect{y}$, 
defined as a curve which starts from $-\infty e^{-i\epsilon}$ and ends at
$-\infty e^{i\epsilon}$ as shown in Fig.~\ref{in-inpath}.
All the external points are also supposed to be placed along this path. 
In case of the wave number
representation, this construction completes 
the definition of the in-in path on the $\eta$-plane. 
If we take the $\eta$-path along $P$, the integral converges 
with the integrand vanishing fast enough in the past.

For the purpose of the present paper, it is more convenient to 
use the position space representation to compute the correlators.  
The vertex integrals involve the spatial integration, too. 
As a starting point, we set the region of the vertex integral 
$\Omega$ to $P\times\mathbf{R}^{D-1}$. 
If we first carry out the spatial integration
before temporal one, the integral would diverge because we then
pick up the contributions from distant spacelike region. 
On the other hand, 
if we integrate first for the time variable and then for the spatial 
ones, the integral is
convergent as we see in Sec.~\ref{sec4}. This means that the integral over
$P\times\mathbf{R}^{D-1}$ is not well-defined as a multiple integral. 

To make the integral to be well-defined as a multiple integral, 
we modify the integral region by deforming the path of the 
$\eta$-integral $P$ \cite{Higuchi:2010xt}.
There are branching points on the $\eta$-plane, which correspond to 
the intersections with the light cones emanating from the external points. 
On the $\eta$-plane for fixed $\vect{y}$, 
$G(X_{i},Y)$ has the same structure of Riemann surface 
as that of $(1-Z_{iY})^{\nu_{i}}$, where $\nu_{i}$ is some
complex number, and 
\beq
\frac{1-Z_{iY}}{2}=\frac{(-\eta+\eta_{i,+})(\eta-\eta_{i,-})}{4(-\eta)(-\eta_{i})}~,
\eeq
where
\beq
\eta_{i,\pm}:=\eta_{i}\pm||\vect{x}_{i}-\vect{y}||~,\qquad(i=1,\cdots,N-1,N)~. 
\eeq
Namely, the integrand has the same structure of Riemann surface as that of
\beq
(-\eta)^{-(D+\sum \nu_{i})}\prod_{i=1}^{N} (-\eta+\eta_{i,+})^{\nu_{i}}(\eta-\eta_{i,-})^{\nu_{i}}~.
\eeq

\begin{figure}[t]
\begin{center}
\includegraphics*[height=5.3cm]{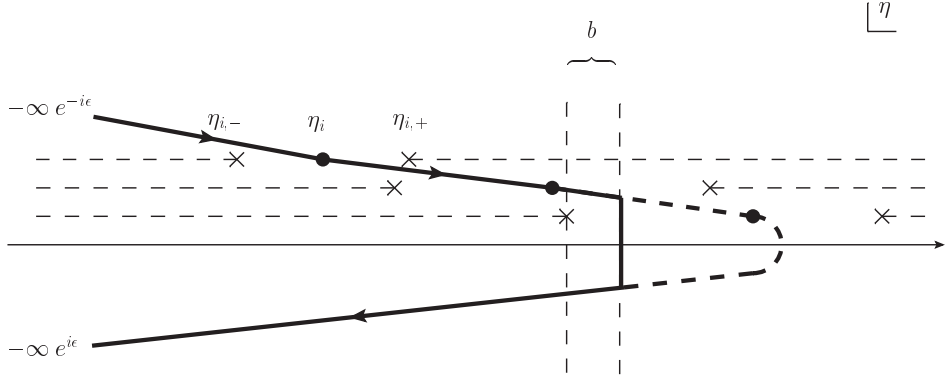}
\caption{\label{in-inpath_deformed}
This figure represents the deformed contour $P_{\svect{y}}$ 
for fixed spatial coordinate $\vect{y}$. The original path $P$ is deformed 
as long as it does not cross the singularities.
}
\end{center}
\end{figure}

The time integration is unchanged even if we deform
the integration contour as long as it does not cross singularities of 
the integrand. 
Thus, we deform the contour $P$ to $P_{\svect{y}}$
such that the maximum value of the real part of $\eta$ on 
$P_{\svect{y}}$ is equal to 
$\max_{i}\{\RR\eta_{i,-}\}+b$ 
where $b$ is a small real positive constant.
(See Fig.~\ref{in-inpath_deformed}.) This deformation
on the $\eta$-plane is significant when the spatial coordinates 
of the vertex is largely separated from those of 
relevant external points. To the contrary, when
$||\vect{x}_{i}-\vect{y}||$ is small for $i$ that realizes 
the maximum among $\RR\eta_{i,-}$, 
the modified contour $P_{\svect{y}}$ is almost identical to 
the original one $P$.

Using this $P_{\svect{y}}$, we define the integration region
\beq
\label{integration region Omega}
\Omega:=\left\{(\eta,\vect{y})\, 
|\, \eta\in P_{\svect{y}},\,\vect{y}\in\mathbf{R}^{D-1} \right\}~,
\eeq
in $\mathbf{C}\times\mathbf{R}^{D-1}$. 
The result of the integral is the same as that is obtained 
by integrating first for time and then for space 
for the original integration region, but we emphasize
that the integral over $\Omega$ is now a multiple
integral.


\subsection{Problems to Be Solved in the Calculations}

Let us return to Eq.~\eqref{single vertex}. Inserting Eq.~\eqref{Barnes Green's function} 
into Eq.~\eqref{single vertex}, we have
\beq
\mathcal{V}_{N}(X_{1},\cdots,X_{N})=\int_{\Omega}dV_{Y}
\int_{\nu_{1}}\cdots\int_{\nu_{N}}
\left[\prod_{i=1}^{N}\Gam(-\nu_{i})\psi(\nu_{i})\right]
\times \left[\prod_{i=1}^{N}\left(\frac{1-Z_{iY}}{2}\right)^{\nu_{i}}\right]~.
\eeq
If we can exchange the order of the integrals, 
$\int_{\Omega}dV_{Y}$ and $\int_{\nu_{1}}\cdots\int_{\nu_{N}}$, 
we are led to calculate the following integral
\beq\label{master integral}
\mathcal{M}(\nu_{1},\cdots,\nu_{N-1},\,\nu_{N})=\int_{\Omega}dV_{Y}\,
\left(\frac{1-Z_{1Y}}{2}\right)^{\nu_{1}}\cdots
\left(\frac{1-Z_{NY}}{2}\right)^{\nu_{N}}~.
\eeq
The first problem is to calculate this integral. This
quantity is shown to have an analytic Mellin-Barnes representation in
Sec.~\ref{sec4}, and hence if this exchange of the order of integration is
allowed, $\mathcal{V}_{N}$ can be represented in an analytic Mellin-Barnes form.
It is not trivial whether this exchange of the order of the
integration is allowed or not. This is the second problems. 
The same problem arises also for arbitrary graphs as for the 
tree level graphs. We will extend our discussion to 
arbitrary graphs in Sec.~\ref{sec5}.

\section{\label{sec4}Computation of the Master Integral}

The goal of this section is to compute the master integral:
\beq
\mathcal{M}(\nu_{1},\cdots,\nu_{n},\,\nu_{N})=\int_{\Omega}dV_{Y}\,
\left(\frac{1-Z_{1Y}}{2}\right)^{\nu_{1}}\cdots
\left(\frac{1-Z_{nY}}{2}\right)^{\nu_{n}}
\left(\frac{1-Z_{NY}}{2}\right)^{\nu_{N}}~,
\label{masterintegral}
\eeq
where we have introduced $n:=N-1$ for convenience, 
\beq
dV_{Y}=\frac{d\eta\,d^{D-1}\! y}{(-\eta)^{D}}~,
\eeq
is the invariant volume, 
and $\Omega$ is defined in Eq.~\eqref{integration region Omega}.

\subsection{\label{Gen. Func. for the Master Integral}Generating Function for the Master Integral}

In order to evaluate the above expression \eqref{masterintegral},
we introduce the following generating function
\beq\label{generating function}
\mathcal{A}(\alpha_{1},\cdots,\alpha_{n}):=
\int_{\Omega}dV_{Y}\,\left(
\sum_{i=1}^{N}\alpha_{i}\frac{1-Z_{iY}}{2}\right)^{\lambda}~, 
\eeq
following Ref.~\cite{Marolf:2010nz}, in which it was used to evaluate 
the master integral on an Euclidean sphere. 
Here 
\beq
\RR\lambda<0~,\quad \alpha_{1}~,~\cdots~,~\alpha_{n}\geq 0~,\quad
\alpha_{N}:=1~, 
\label{lambdaconditions}
\eeq
are assumed. 

In this subsection we establish the relation between the 
generating function 
and the master integral. 
Formally, 
in the same way as in the Euclidean case discussed in Ref.~\cite{Marolf:2010nz},
the generating function \eqref{generating function} seems to be related 
to the master integral
\eqref{master integral} also in the present case as follows:

\noindent
[Step 1.] We first apply Eq.~\eqref{Barnes formula2} to 
the integrand of \eqref{generating function} to obtain
\beq\label{Step1desu}
\mathcal{A}(\alpha_{1},\cdots,\alpha_{n})&=&
\int_{\Omega}dV_{Y}\, \frac{1}{\Gam(-\lambda)}
\int_{u_{1}}(\alpha_{1})^{u_{1}}\cdots\int_{u_{n}}(\alpha_{n})^{u_{n}}\nonumber\\
&&\quad \times\Gam\left[-u_{1},\,\cdots,\, -u_{n},\, -u_{N}\right]
\left(\frac{1-Z_{1X}}{2}\right)^{u_{1}}\cdots \left(\frac{1-Z_{NX}}{2}\right)^{u_{N}}~,
\eeq
where
\beq
u_{N}:=\lambda-\sum_{i=1}^{n}u_{i}~.
\label{defuN}
\eeq

\noindent
[Step 2.] Next, we exchange the order of the integration, 
$\int_{\Omega} dV_{Y}$ and $\int_{u_{1}} \cdots \int_{u_{n}}$, to have 
\beq\label{Barnes integral form of A}
\mathcal{A}(\alpha_{1},\,\cdots,\, \alpha_{n})
&=&\frac{1}{\Gam(-\lambda)}
\int_{u_{1}}(\alpha_{1})^{u_{1}}\cdots\int_{u_{n}}(\alpha_{n})^{u_{n}}\nonumber\\
&&\quad \times\Gam\left[-u_{1},\,\cdots,\, 
-u_{n},\, -u_{N}\right]\mathcal{M}\left(u_{1},\,\cdots,\, u_{n},\, u_{N}\right)~.
\eeq
Thus, the Mellin transform of $\mathcal{A}$ gives $\mathcal{M}$. 

However, we have to prove that [Step 1.] and [Step 2.] are
indeed possible, which is the goal of this subsection. 
In particular, [Step 2.] requires that the integral over $\Omega$ 
is a multiple integral. The convergence of the integral 
is rather obvious when we consider the corresponding integral 
over a compact Euclidean sphere, while it is not 
in the present case where the integration region is non-compact. 
In this subsection,
we assume, for a technical reason, that the time coordinates 
of all external points lie 
on the real Lorentzian section, i.e. $\eta_{i}\in
\mathbf{R}_{-},\,\vect{y}_{i}\in\mathbf{R}^{D-1}$, and 
furthermore, that any pairs of them are mutually 
spacelike separated.

\begin{figure}[t]
\begin{center}
\includegraphics*[height=4cm]{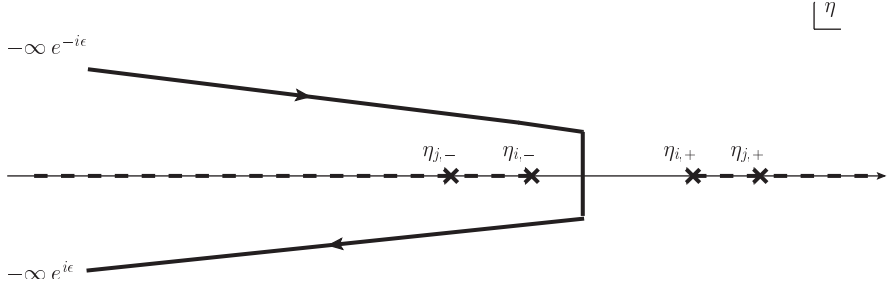}
\caption{
\label{in-inpath deformed real Lorentzian}
A figure representing the in-in path $P_{\svect{y}}$ for the external points 
which lie on the real Lorentzian section and are mutually spacelike separated. 
The crosses represent the branching points and the dotted lines the branch cuts.
}
\end{center}
\end{figure}

Since the definition of the in-in path described in Sec.~\ref{sec3.1} 
requires the external points to lie along the in-in path and 
therefore their time coordinates are complex in general, we need 
some explanations of the in-in path for this configuration.
The path is defined on the $\eta$-plane by taking the limit $\II\eta_{i}\to 0$
in $P_{\svect{y}}$ introduced in Sec.~\ref{sec3.1}.
It seems that the path in this limit must, at least partly, 
lie on the $\eta$-real axis.
However, since the 
external points are mutually spacelike, 
the branch cuts, lying on the $\eta$-real axis,  
do not cover the whole $\eta$-real axis.
Therefore, the limit can be taken without
the pass $P_{\svect{y}}$ crossing the branch cuts,
and hence the in-in path in this limit is simply
a contour going from $-\infty\,e^{-i\epsilon}$ to $-\infty\, e^{i\epsilon}$
as shown in Fig.~\ref{in-inpath deformed real Lorentzian}.

\begin{figure}[t]
\begin{center}
\includegraphics*[height=5cm]{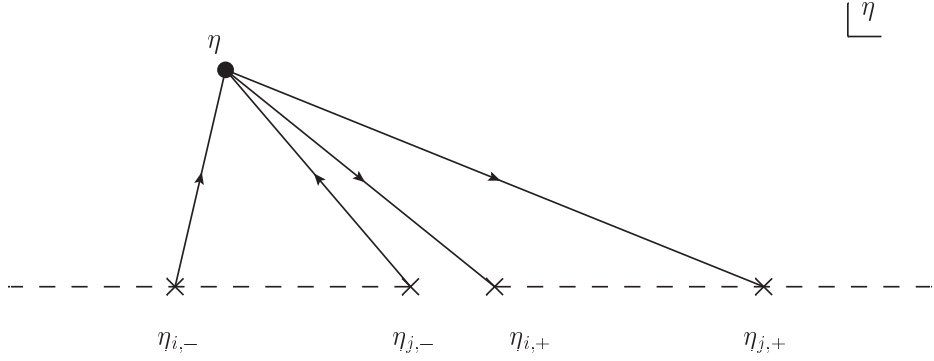}
\caption{\label{arguments}The dot represents the time coordinate of a vertex on $P_{\svect{y}}$.
The remainder of the summation of the arguments of two vectors relevant to the subscript $i$ and that relevant to the subscript $j$ gives $|\arg(1-Z_{iY})-\arg(1-Z_{jY})\,|$ as in \eqref{argumentsformula}.
}
\end{center}
\end{figure}

Proof of [Step 1.]: 
Note that the following inequalities hold for arbitrary $Y\in \Omega$:
\beq\label{condition for Barnes formula2}
|\arg(1-Z_{iY})-\arg(1-Z_{jY})\,|<\pi~,~
\quad(i,j=1,2,\cdots,N)~.
\eeq
In fact, $\arg(1-Z_{iY})$ is given by
\beq
\arg(1-Z_{iY})=\arg(-\eta+\eta_{i,+})+
\arg(\eta-\eta_{i,-})-\arg(-\eta)-\arg(-\eta_{i})~,
\eeq
and then, noticing that $\arg(-\eta_{i})=0$ 
since all the external points are on the real Lorentzian section, we have
\beq
\label{argumentsformula}
&&|\arg(1-Z_{iY})-\arg(1-Z_{jY})\,|\nonumber\\
&&=|\arg(-\eta+\eta_{i,+})+\arg(\eta-\eta_{i,-})-\arg(-\eta+\eta_{j,+})-\arg(\eta-\eta_{j,-})\,|~.
\eeq
This quantity is less than $\pi$ for any
$(\eta,\vect{y})\in\Omega$. (See Fig.~\ref{arguments}.) 
The inequality \eqref{condition for Barnes
formula2} is the sufficient condition that the formula 
\eqref{Barnes formula2} can
be applied to the integrand of Eq.~\eqref{generating function}.
For the later purpose, we modify the integration path
$P_{\svect{y}}$ 
as such that satisfies 
\begin{equation}
 |\arg(1-Z_{iY})-\arg(1-Z_{jY})|<\pi -\delta~, 
\label{argbound}
\end{equation}
for any $i$ and $j$ with a small positive number $\delta$.  
This can be achieved easily. Because
$|\arg(1-Z_{iY})-\arg(1-Z_{jY})|$ is close to $\pi$ only in the
small region surrounding the interval $(\eta_{i,-},\eta_{j,-})$ or $(\eta_{i,+},\eta_{j,+})$, 
the path can be chosen to avoid this region.

Proof of [Step 2.]: 
We denote the integration paths for $u_{1},\cdots,u_{n}$ as
$C_{1},\cdots,C_{n}$, respectively, and define
$C:=C_{1}\times\cdots\times C_{n}$. 
The sufficient condition to allow to
exchange the order of the integration,
$\int_{\Omega}dV_{Y}$ and $\int_{C}\prod_{i=1}^{n}du_{i}/2\pi i$, is
that the integral is absolutely convergent (Fubini's
theorem). In the present case, we should examine 
the following integral
\beq\label{absolute convergence for master integral}
&&\frac{1}{|\Gam(-\lambda)|}\int_{C}\prod_{k=1}^{n}\left|\frac{du_{k}}{2\pi i}\right|
\left|(\alpha_{1})^{u_{1}}\cdots(\alpha_{n})^{u_{n}}\right|
\left|\Gam\left[-u_{1},\,\cdots,\, -u_{n},-u_{N}\right]\right|\nonumber\\
&&\qquad\times\left[\int_{\Omega}|dV_{Y}|\,
\left|\left(\frac{1-Z_{1Y}}{2}\right)^{u_{1}}\right|\cdots \left|\left(\frac{1-Z_{NY}}{2}\right)^{u_{N}}\right|\,
\right]~,
\eeq
where
\beq
|dV_{Y}|=\frac{|d\eta|d^{D-1}\! y}{|-\eta|^{D}}~.
\eeq
If this integral is finite, then we can justify 
the exchange of the order of integrals in [Step 2.]. 

To show this, 
we focus on the integrand of the $\Omega$ integral 
in the large brackets in 
Eq.~\eqref{absolute convergence for master integral}
for fixed $u_{1},\cdots,u_{n}$:
\beq\label{integrand for Omega integration}
\left|\left(\frac{1-Z_{1Y}}{2}\right)^{u_{1}}\right|\cdots
\left|\left(\frac{1-Z_{NY}}{2}\right)^{u_{N}}\right|~. 
\eeq
Notice that
\beq\label{transform1}
\left|(1-Z_{iY})^{u_{i}}\right|=|1-Z_{iY}|^{\RR u_{i}}
\exp\left[-\arg(1-Z_{iY})\II u_{i}\right]~.
\eeq
Along the integration path of $u_i$ parallel to the imaginary axis, 
$\II u_i$ 
varies while $\RR u_i$ is fixed. 
Taking into account that $u_N$ includes $u_i$ as given in Eq.~\eqref{defuN}, 
the part depending on $\II u_i$ in 
Eq.~\eqref{integrand for Omega integration} is factored out as 
\begin{equation}
 \exp\left[\left\{\arg(1-Z_{NY})-\arg(1-Z_{iY})\right\}
       \II u_{i}\right]~. 
\end{equation}
Since $\left|\arg(1-Z_{NY})-\arg(1-Z_{iY})\right|$ is bounded 
as shown in Eq.~\eqref{argbound}, this factor is bounded from above by 
$
 \exp\left[(\pi -\delta)|\II u_{i}|\right].
$
Therefore, noticing that $\alpha_i$ is real positive number, 
we find that 
\begin{eqnarray}
[\mbox{Eq.}~\eqref{absolute convergence for master integral} ]
&<&
\frac{1}{|\Gam(-\lambda)|}
\left|(\alpha_{1})^{\RR u_{1}}\cdots(\alpha_{n})^{\RR u_{n}}\right|
\int_{\Omega}|dV_{Y}|\,
\left|\left(\frac{1-Z_{1Y}}{2}\right)\right|^{\RR u_{1}}\cdots
\left|\left(\frac{1-Z_{NY}}{2}\right)\right|^{\RR u_{N}}\, 
\cr &&\qquad\times
\int_{C}\prod_{k=1}^{n}\left|\frac{du_{k}}{2\pi i}\right|
\left|\Gam\left[-u_{1},\,\cdots,\, -u_{n},-u_{N}\right]\right|
 e^{(\pi -\delta)\left(|\II u_{1}|+\cdots +|\II u_{n}|\right)}~.
\end{eqnarray}
Since 
$|\Gam(x+iy)|\approx
(2\pi)^{1/2}e^{-\pi|y|/2}|y|^{x-1/2}\,(|y|\to +\infty)$, 
$u_k$ integrals in the second line 
in the last expression are convergent. 
Therefore, our remaining task is to show that the volume 
integral 
\beq
\label{absolute convergence for Omega integration}
\int_{\Omega}|dV_{Y}|\,
\left|\left(\frac{1-Z_{1Y}}{2}\right)\right|^{\RR u_{1}}
\cdots
\left|\left(\frac{1-Z_{NY}}{2}\right)\right|^{\RR u_{N}}
,
\eeq
is also finite. 


For this purpose, we first introduce a representative point $X_{0}$ 
with coordinates in the flat chart defined by 
\beq
(\eta_{0},\vect{x}_{0}):=\sum_{i=1}^{N} p_{i}(\eta_{i},\vect{x}_{i})~,
\qquad(p_{i}\geq 0,\,\sum p_{i}=1)~,
\eeq
and a domain $D_{0}$ far from $X_{0}$ in terms of the invariant distance 
by 
\beq\label{definition of D_{0}}
D_{0}:=\left\{Y\,|\, |Z_{0Y}|>Z_{0}\right\}\cap \Omega~.
\eeq
Note that if we take $Z_{0}$ to be sufficiently large, we see that
\beq\label{large Z0}
|Z_{iY}|\geq \mathrm{const.}\times |Z_{0Y}|~,\qquad (Y\in D_{0},\, i=1,2,\cdots,N)~.
\eeq
We divide the region $\Omega$ into $(\Omega\backslash D_{0})$ and
$D_{0}$, and evaluate each contribution to \eqref{absolute convergence
for Omega integration} separately.

\quad\\
\noindent
\underline{(i) Integral over $\Omega\backslash D_{0}$:}

\begin{figure}[t]
\begin{center}
\includegraphics*[height=7cm]{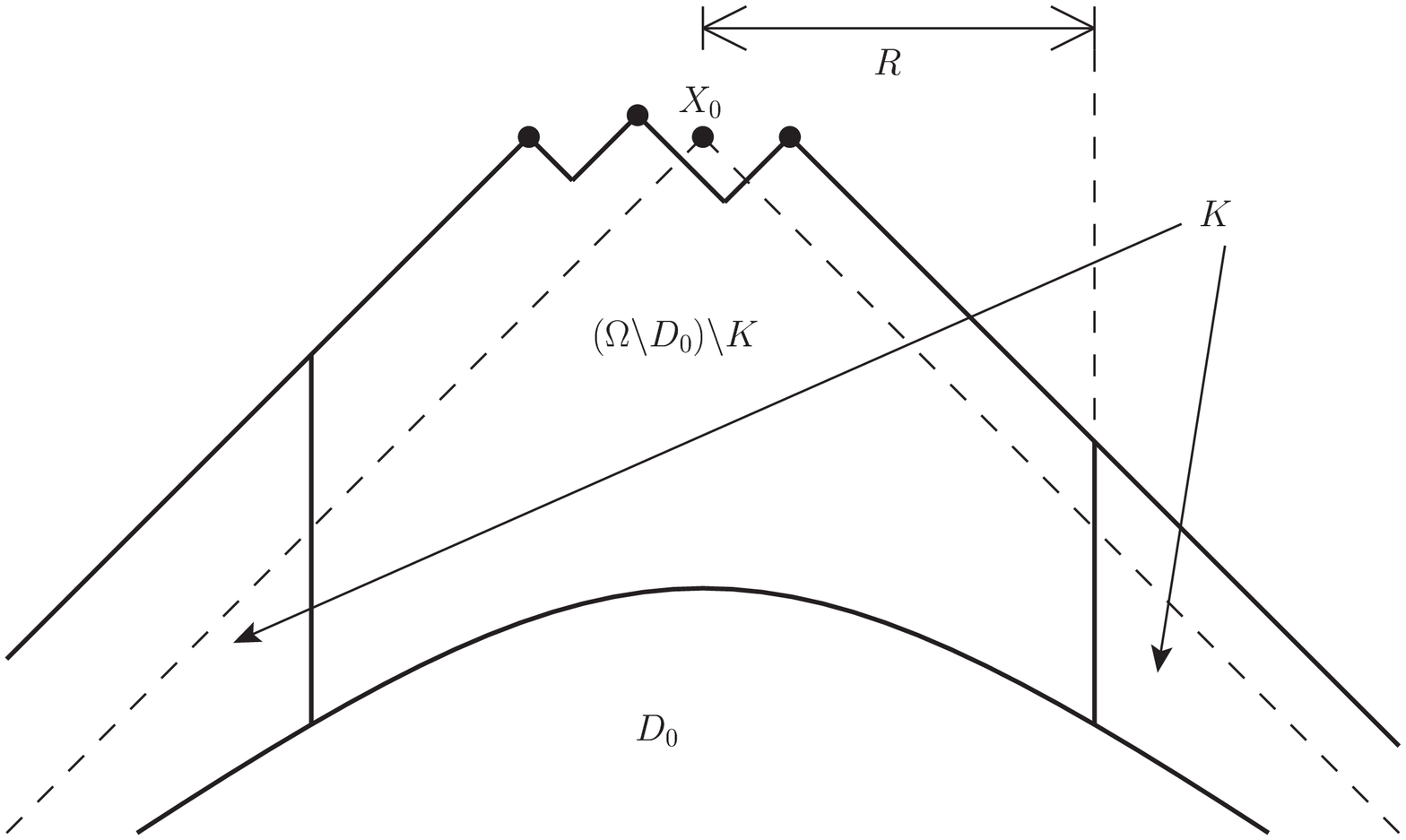}
\caption{\label{Omega}This figure is a schematic of how we divide the integration region $\Omega$. There are $D_{0}$, $K$ and $(\Omega\backslash D_{0})\backslash K$. The dots except for $X_{0}$ represent the external points. The dashed lines represent schematically the ``past light cone of $X_{0}$.''}
\end{center}
\end{figure}

We further divide $\Omega\backslash D_{0}$ into $K$ defined by 
\beq
K:=\left\{(\eta,\vect{y})\in \Omega\backslash D_{0}\,|\, ||\vect{x}_{0}-\vect{y}||>R  \right\}~,
\eeq
and its complement $(\Omega\backslash D_{0})\backslash K$.
$R$ is set large enough 
for $K$ not to include any external points. 
(See Fig.~\ref{Omega}.)

\quad\\
\noindent
\underline{(i-a) Integral over $(\Omega\backslash D_{0})\backslash K$:}

The region $(\Omega\backslash D_{0})\backslash K$ 
is compact but 
it contains the coincidence points $(\eta,\vect{y})=(\eta_{i},\vect{x}_{i})$ 
at which the integrand of 
\eqref{absolute convergence for Omega integration} diverges.
Since $\vect{y}\approx
\vect{x}_i$ around them, the path $P_{\svect{y}}$ is identical to 
the original one $P$, and hence
$\eta=\eta_{i}+i\delta\eta$ with real $\delta\eta$.  
Then, we have
\beq
|1-Z_{iY}|^{\RR u_{i}}\approx ((\delta\eta)^{2}+||\vect{x}_{i}-\vect{y}||^{2})^{\RR u_{i}}
\eeq
around a point $(\eta_{i},\vect{x}_{i})$, which shows that
\eqref{absolute convergence for Omega integration} is finite as long as
we choose the integration path of $u_i$ to satisfy 
\beq\label{conditions for uis}
\RR u_{i}
>-D/2~,\quad(i=1,2,\cdots, n, N)~,
\eeq
which does not conflict with [Step 1.]. 
Recall that the fundamental strip of Eq.~\eqref{Step1desu} contains 
the paths with $\RR u_{i}$ for all $i$ being infinitesimally small negative constants. 

\quad\\
\noindent
\underline{(i-b) Integral over $K$:}

We first see that, 
for $Y\in K$, $|1-Z_{iY}|$ is bounded both from below and from above 
by positive constants. 
Recall that the $\eta$-path $P_{\svect{y}}$ is
defined by deforming $P$ not to touch 
$\eta_{i,-}$ except for the case with $\vect{y}\approx \vect{x}_i$, 
which occurs in $(\Omega\backslash D_{0})\backslash K$. 
Therefore, $|1-Z_{iY}|$ does not vanish, bounded
from below by some constant $c_- (\,>0)$. 
It is also easy to show that $|1-Z_{iY}|$ is bounded from above 
by some constant $c_+$. If $|1-Z_{iY}|$ is sufficiently large, 
$Z_{0Y}$ will be larger than $Z_{0}$. 
Then, by the definition of $K$, $Y$ is not included in $K$. 
Thus, we conclude that for some positive constants $c_{\pm}$,
\beq
c_{-}<|1-Z_{iY}|<c_{+}~,\quad (Y\in K)~.
\eeq
Furthermore, one can claim that the volume of the region $K$ is 
finite, i.e.,  
\beq
\int_{K}|dV_{Y}|<+\infty~.
\eeq
In showing this, the non-trivial point is that the region $K$ 
extends to infinitely large $||\vect{y}||$. However, the region 
of the $\eta$-integral is confined to the interval  
\beq
\label{region covering K}
\eta_{i(\svect{y}),-}-b'\leq\RR
\eta\leq \eta_{i(\svect{y}),-}+b~,
\eeq
where $b$ is the same constant used in defining the path
$P_{\svect{y}}$ and $i(\vect{y})$ is the label of the external point 
such that $\eta_{i(\svect{y}),-} > \eta_{k,-}$ for all $k\ne i(\vect{y})$. 
Here the point is that one can choose 
a large positive constant $b'$ to be independent 
of $\vect{y}$. In fact, the
invariant distance between $X_{0}:=(\eta_{0},\vect{x}_{0})$, and
the point corresponding to the above lower bound
$Y_{\mathrm{bdry}}:=(\eta_{i(\svect{y}),-}-b',\vect{y})=(\eta_{i(\svect{y})}-||\vect{x}_{i(\svect{y})}-\vect{y}||-b',\vect{y})$
is evaluated as 
\beq
|1-Z(X_{0},Y_{\mathrm{bdry}})|&=&\left|\frac{(\eta_{0}-\eta_{i(\svect{y})}+||\vect{x}_{i(\svect{y})}-\vect{y}||+b')^{2}-||\vect{x}_{i(\svect{y})}-\vect{y}||^{2}}
{2\eta_{0}(\eta_{i(\svect{y})}-||\vect{x}_{i(\svect{y})}-\vect{y}||-b')}\right|\nonumber\\
&\gtrsim & \frac{b'}{|-\eta_{0}|}~,~
\qquad (||\vect{x}_{i(\svect{y})}-\vect{y}||+b'\to +\infty)~.
\eeq
In the last inequality we assumed 
$||\vect{x}_{i(\svect{y})}-\vect{y}||+b'\to +\infty$ but this should be 
a good approximation in the region $K$. 
Therefore, if $b'/(\,|\!-\eta_{0}|)$ is taken sufficiently large compared 
with $Z_{0}$, the above range of $\eta$ covers 
the whole region of $K$. 
Thus, the volume $\int_{K}|dV_{Y}|$ is bounded by 
\beq
\int_{K}|dV_{Y}|< 
c_1
\int_{||\svect{x}_{0}-\svect{y}||>R}
d^{D-1}y
\int_{\eta_{i(\svect{y}),-}-b'}^{\eta_{i(\svect{y}),-}+b}{|d\eta| \over |\eta|^D}
< c_2 
\int_{||\svect{x}_{0}-\svect{y}||>R} {d^{D-1}y\over ||\vect{x}_{i(\svect{y})}-\vect{y}||^D}<+\infty~,
\eeq
where $c_1$ and $c_2$ are some appropriately chosen constants of $O(1)$.
In the second inequality we used $|\eta|>|\eta_{i(\svect{y}),-}+b|\approx
||\vect{x}_{i(\svect{y})}-\vect{y}||$. 
Therefore, the integral over $K$ is proven to be finite.

\quad\\
\noindent
\underline{(ii) Integral over $D_{0}$:}

We next proceed to the integral over $D_{0}$. 
Using Eq.~\eqref{large Z0}, one can easily bound the volume 
integral of our current concern from above as
\beq
\int_{D_{0}}|dV_{Y}|\,
\left|\left(\frac{1-Z_{1Y}}{2}\right)\right|^{\RR u_{1}}\cdots
\left|\left(\frac{1-Z_{NY}}{2}\right)\right|^{\RR u_{N}}
 < c_3 \times 2^{-\lambda}\int_{D_{0}}|dV_{Y}|\,|Z_{0Y}|^{\RR \lambda}~.
\label{expressiontoevaluate}
\eeq
where $c_3$ is a constant of $O(1)$, and we have used the relation 
$\sum^{N} u_{i}=\lambda$.

In order to show this integral is finite, 
we use $Z_{0Y}$ as a time coordinate instead of $\eta$,
which leads the integration measure to transform as
\beq
d\eta\, d^{D-1}\! y=\left(1-\frac{\eta_{0}Z_{0Y}}{\sqrt{||\vect{x}_{0}-\vect{y}||^{2}
+\eta_{0}^{2}(Z_{0Y}^{2}-1)}}\right)\eta_{0}d(Z_{0Y})d^{D-1}\! y~.
\eeq
We substitute this into the right hand side of \eqref{expressiontoevaluate}.
Approximating $Z_{0Y}^{2}-1\approx Z^{2}_{0Y}$ 
and introducing $x:=y/(-\eta_{0}Z_{0Y})$, 
we find that the integral is finite as 
\beq
\int_{D_{0}}|dV_{Y}|\, |Z_{0Y}|^{\RR \lambda} < 
c_4 \int_{Z_0}^{+\infty} \frac{dZ}{Z^{1-\RR\lambda}}
\int_{0}^{+\infty} dx\,{1\over x/\sqrt{1+x^{2}}}
\left(
{x\over (1+\sqrt{1+x^{2}})}\right)^{D-1} < +\infty ~, 
\eeq
where again $c_4$ is a constant of $O(1)$. 

\quad\\
\noindent
\underline{(iii) Summary:}

We have shown in this subsection that the integral \eqref{absolute
convergence for master integral} is indeed finite when the external
points $X_{1},\cdots,X_{n}$ and $X_{N}$ lie on the real Lorentzian 
section and are mutually in spacelike separation,  
as long as the integration contours for $u_{1},\cdots, u_{n}$ 
satisfy the additional conditions \eqref{lambdaconditions} and 
\eqref{conditions for uis}: 
\beq
&&\RR\lambda=\RR\sum^{N}_{i}u_{i}<0~,\nonumber\\
&&\RR u_{1}>-\frac{D}{2},~\cdots~, \quad \RR u_{n}>-\frac{D}{2},\quad \RR u_{N}>-\frac{D}{2}~.
\label{finiteMasterIntegral}
\eeq

Then, the order of two integrals $\int_{\Omega}dV_{Y}$ and
$\int_{C}\prod_{i=1}^{n}du_{i}/2\pi i$ in Eq.~\eqref{Step1desu} are
exchangeable, which implies that the master integral $\mathcal{M}$ is
given by the repeated Mellin transform of $\mathcal{A}$. Furthermore,
under these conditions $\mathcal{A}(\alpha_{1},\cdots, \alpha_{n})$ is
finite and thus from Eq.~\eqref{Barnes integral form of A} the master
integral $\mathcal{M}(u_{1},\cdots,u_{n},u_{N})$ is also finite. That
is, the master integral $\mathcal{M}(u_{1},\cdots, u_{n}, u_{N})$ is finite
when the external points are in the real Lorentzian section and are
mutually in spacelike separation, with the conditions 
\eqref{finiteMasterIntegral} satisfied. 
The analytic expression for $\mathcal{M}$ is given in the succeeding 
subsection, where the conditions on the external points 
are relaxed.

\subsection{\label{Cal. of Gen. Func.}Calculation of the Generating Function}

We now proceed to compute $\mathcal{A}$ and hence $\mathcal{M}$,  
to show its equivalence to the analytic continuation of the 
Euclidean correlators. Again in this subsection we first 
assume that all the external points $X_{i}$ lie on the real
Lorentzian section and that they are mutually in spacelike separation. 
After that, we show that the time coordinates of the external points
$\eta_{i}$ in the obtained expression for $\mathcal{M}$ 
can be analytically continued to any point on the in-in path. 

The expression for $\mathcal{A}$ 
given in Eq.~\eqref{generating function} can be transformed into 
\beq
\mathcal{A}(\alpha_{1},\cdots,\alpha_{n})&=&\int_{\mathbf{R}^{D-1}}d^{D-1}\! y\int_{P_{\svect{y}}}\frac{d\eta}{(-\eta)^{D}}\,
2^{-\lambda}\left(\sum_{i=1}^{N}\alpha_{i}-V\cdot Y\right)^{\lambda}~,
\eeq
where $V\cdot Y$ is an inner product of $V$ and $Y$ 
with respect to $(D+1)$-dimensional Minkowski metric 
and 
\beq
V=\sum_{i=1}^{N}\alpha_{i}X_{i}~.
\eeq

Notice that
\beq
V^{0}+V^{D}=\sum_{i=1}^{N}{\alpha_{i}\over -\eta_{i}}>0~,\quad
\vect{V}=\sum_{i=1}^{N}\frac{\alpha_{i}}{-\eta_{i}}\vect{x}_{i}~. 
\eeq
Setting
\beq
&&\tau:=\frac{V^{0}+V^{D}}{2}\eta~,
\quad
\vect{R}:=(V^{0}+V^{D})\vect{x}-\vect{V}=\sum_{i=1}^{N}\frac{\alpha_{i}}{-\eta_{i}}(\vect{y}-\vect{x}_{i})~,
\eeq
$V\cdot Y$ can be expressed as 
\beq
V\cdot Y
=-\tau+\frac{\vect{R}^{2}-V\cdot V}{4}\frac{1}{\tau}~,
\eeq
where
\beq
V\cdot V:=\eta_{ab}V^{a}V^{b}=\left(\sum_{i=1}^{N}\alpha_{i}\right)^{2}+2\sum_{i<j}^{N}\alpha_{i}\alpha_{j}(Z_{ij}-1)~.
\eeq
Thus, we obtain
\beq\label{tochu1}
\mathcal{A}(\alpha_{1},\cdots,\alpha_{n})&=&\int_{\mathbf{R}^{D-1}}d^{D-1}\! y\int_{P_{\svect{y}}}\frac{d\eta}{(-\eta)^{D}}\,
2^{-\lambda}\left(\sum_{i=1}^{N}\alpha_{i}-V\cdot Y\right)^{\lambda}\nonumber\\
&=&2^{-\lambda}
\int_{\mathbf{R}^{D-1}}d^{D-1}\! y\left(\frac{V^{0}+V^{D}}{2}\right)^{D-1}\int_{P'_{\svect{y}}}d\tau\,(-\tau)^{-D}\left[
\sum_{i=1}^{N}\alpha_{i}+\tau-\frac{\vect{R}^{2}-V\cdot V}{4}\frac{1}{\tau}
\right]^{\lambda}
\nonumber\\
&=&
2\pi i\times 2^{-D+1-\lambda}
\int_{\mathbf{R}^{D-1}}d^{D-1}\! R\int_{P'_{\svect{y}}}\frac{d\tau}{2\pi i}\, (-\tau)^{-D-\lambda}(-\tau+\tau_{+})^{\lambda}(\tau-\tau_{-})^{\lambda}~.
\eeq
where $P'_{\svect{y}}$ is the scale-transformed path of $P_{\svect{y}}$
by a factor of $(V^{0}+V^{D})/2$, 
and 
\beq
\label{definition of C and G}
\tau_{\pm}:=\frac{1}{2}\left\{-F\pm\sqrt{\vect{R}^{2}+J^{2}}\right\}~,\quad
F:=\sum_{i=1}^{N}\alpha_{i}~,\quad
J^{2}:=2\sum_{i<j}^{N}\alpha_{i}\alpha_{j}(1-Z_{ij})~.
\eeq
As is mentioned at the beginning of this subsection, we have assumed
that the external points are all mutually in spacelike separation, 
so that
\beq
J^{2}>0~,\quad \tau_{\pm}\in \mathbf{R}~.
\eeq

\begin{figure}[t]
\begin{center}
\includegraphics*[height=5cm]{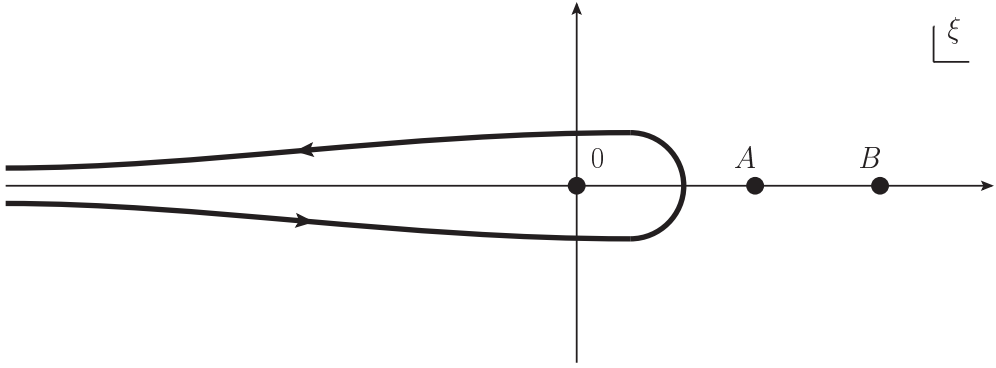}
\caption{\label{Cpath}The path $C$ in the integral \eqref{formula1}.}
\end{center}
\end{figure}

Changing the integration variable further to 
$\xi:=\tau-\tau_{-}$, we obtain 
\beq\label{tochu2}
\mathcal{A}(\alpha_{1},\cdots,\alpha_{n})=-2\pi i\times 2^{-D+1-\lambda}
\int_{\mathbf{R}^{D-1}}d^{D-1}\! R\int_{C}\frac{d\xi}{2\pi i}\, (A-\xi)^{-D-\lambda}(B-\xi)^{\lambda}\xi^{\lambda}~,
\eeq
where 
\beq
A:=-\tau_{-}=\frac{1}{2}\left\{F+\sqrt{\vect{R}^{2}+J^{2}}\right\}~,\quad
B:=\tau_{+}-\tau_{-}=\sqrt{\vect{R}^{2}+J^{2}}~,
\eeq
are both positive, and 
$C$ is the integration contour shown in Fig.~\ref{Cpath}, 
which corresponds to $P'_{\svect{y}}$ but the direction is reversed.
(See Fig.~\ref{in-inpath deformed real Lorentzian} as a reference.)
Notice that the path $C$ does not have to respect the $i\epsilon$ 
prescription here, because this integral is convergent without relying on 
the $i\epsilon$ regulator. 

In order to compute this integral, we 
use the formula 
\beq\label{formula1}
&&\int_{C}\frac{d\xi}{2\pi i}\,(A-\xi)^{\alpha}(B-\xi)^{\beta}\xi^{\gamma}\nonumber\\
&&\qquad =\int_{\mu}\Gam\left[
\begin{matrix}
-\alpha+\mu,\, -(\alpha+\beta+\gamma+1)+\mu,\, -\mu,\, \alpha+\gamma+1-\mu\\
-\alpha,\, -\beta,\, -\gamma,\, \gamma+1
\end{matrix}
\right]
A^{\mu}B^{\alpha+\beta+\gamma+1-\mu}~,
\eeq
which is valid for $\RR(\alpha+\beta+\gamma+1)<0$. 

It is not difficult to verify \eqref{formula1}. We denote this integral
as $I(\alpha,\,\beta,\,\gamma)$. When $\RR\gamma>-1$, the path $C$ can
be contracted to the forward and backward paths along 
the negative real axis. Noticing that only the argument of
$\xi^{\gamma}$ changes between these two paths, one can transform
$I(\alpha,\,\beta,\,\gamma)$ as 
\beq
I(\alpha,\,\beta,\,\gamma)
&=&\frac{1}{2\pi i}\int_{+\infty}^{0}(-dx)e^{-i\pi\gamma}x^{\gamma}(x+A)^{\alpha}(x+B)^{\beta}
+\frac{1}{2\pi i}\int_{0}^{+\infty}(-dx)e^{i\pi\gamma}x^{\gamma}\cdots\nonumber\\
&=&\frac{1}{\Gam[-\gamma,\,\gamma+1]}\int_{0}^{+\infty}dx\,(x+A)^{\alpha}(x+B)^{\beta}x^{\gamma}~.
\label{I1}
\eeq
Next, we expand $(x+B)^{\beta}$, using Eq.~\eqref{Barnes formula1}, as
\beq
(x+B)^{\beta}=\int_{\mu}\Gam\left[
\begin{matrix}
-\beta+\mu,\, -\mu\\
-\beta
\end{matrix}
\right]x^{\mu}B^{\beta-\mu}~,\qquad(\RR\beta<\RR\mu<0)~.
\eeq
Substituting this into Eq.~\eqref{I1}, 
we carry out $x$ integral first to obtain 
\beq
I(\alpha,\,\beta,\,\gamma)=\int_{\mu}\Gam\left[
\begin{matrix}
-\beta+\mu,\, -\mu,\, \gamma+1+\mu,\, -\alpha-\gamma-1-\mu\\
-\gamma,\, \gamma+1,\, -\beta,\, -\alpha
\end{matrix}
\right]A^{\alpha+\gamma+1+\mu}B^{\beta-\mu}~.
\eeq
Of course, the convergence of $x$ integral imposes a condition
$\RR(\alpha+\gamma+1+\mu)<0$. This is in fact satisfied because we can
set $\RR\mu$ arbitrarily close to $\RR\beta$ as long as
$\RR\beta<\RR\mu(<0)$ is maintained. If we change the integration 
variable from $\mu\to\mu-\alpha-\gamma-1$, 
we obtain the expression \eqref{formula1}.

Finally, we remove the restriction $\RR\gamma>-1$. In fact, the
integrand is analytic for $\gamma$ and the $\xi$ integration is
uniformly convergent for $\gamma$, as long as
$\RR(\alpha+\beta+\gamma+1)<0$. Therefore, the integral is analytic for
$\gamma$, which enables us to remove the restriction $\RR\gamma>-1$ by analytic
continuation.

Substituting Eq.~\eqref{formula1} into Eq.~\eqref{tochu2}, we find
\beq\label{generating function tochu1}
\mathcal{A}(\alpha_{1},\cdots,\alpha_{n})
&\!\!=&\!\!-2\pi i\times 2^{-D-\lambda+1}\int_{\mathbf{R}^{D-1}}d^{D-1}\! R\,
\int_{\mu}\Gam\left[
\begin{matrix}
D+\lambda+\mu,\, D-\lambda-1+\mu,\, -\mu,\, -D+1-\mu\\
D+\lambda,\, -\lambda,\, -\lambda,\, \lambda+1
\end{matrix}
\right]\nonumber\\
&&
\qquad\times 2^{-\mu}\left[F+\sqrt{\vect{R}^{2}+{J}^{2}}\right]^{\mu}(\vect{R}^{2}+J^{2})^{\frac{-D+\lambda+1-\mu}{2}}~.
\eeq
We next carry out $\vect{R}$ integration, 
using the formula 
\beq
\label{formula2}
&&\int_{\mathbf{R}^{D-1}}d^{D-1}\! R\,\left[\vect{R}^{2}+J^{2}\right]^{\nu/2}\left[F+\sqrt{\vect{R}^{2}+J^{2}}\right]^{\mu}\nonumber\\
&&\qquad=
\pi^{\frac{D-1}{2}}\int_{\kappa}\Gam\left[
\begin{matrix}
-\mu+\kappa,\, -\kappa,\, -\frac{\kappa+\nu+D-1}{2}\\
-\mu,\, -\frac{\kappa+\nu}{2}
\end{matrix}
\right]
F^{\mu-\kappa}J^{\nu+\kappa+D-1}~, 
\eeq
which is valid when $\RR(\mu+\nu+D-1)<0$. 
The idea of the proof of the above formula is not so different from that of 
the formula \eqref{formula1}. One applies Eq.~\eqref{Barnes formula1} to 
$\left[F+\sqrt{\vect{R}^{2}+J^{2}}\right]^{\mu}$ to obtain
\beq
\left[F+\sqrt{\vect{R}^{2}+J^{2}}\right]^{\mu}=\int_{\kappa}\Gam\left[
\begin{matrix}
-\mu+\kappa,\, -\kappa\\
-\mu
\end{matrix}
\right]
F^{\mu-\kappa}(\vect{R}^{2}+J^{2})^{\kappa/2}~,\qquad(\RR\mu<\RR\kappa<0)~.
\eeq
Substituting this into the left hand side of Eq.~\eqref{formula2}, 
we obtain 
\beq
&&\int_{\mathbf{R}^{D-1}}d^{D-1}\! R\,\left[\vect{R}^{2}+J^{2}\right]^{\nu/2}\left[F+\sqrt{\vect{R}^{2}+J^{2}}\right]^{\mu}\nonumber\\
&&\qquad =\Omega_{D-2}\int_{\kappa}\Gam\left[
\begin{matrix}
-\mu+\kappa,\, -\kappa\\
-\mu
\end{matrix}
\right]
F^{\mu-\kappa}J^{\nu+\kappa+D-1}\times\frac{1}{2}\int_{0}^{+\infty}d\Xi\, \Xi^{\frac{D-3}{2}}(1+\Xi)^{\frac{\nu+\kappa}{2}}~,
\eeq
where we have introduced a new integration variable $\Xi:=(||\vect{R}||/J)^{2}$,
and 
\beq
\Omega_{D-2}=\frac{2\pi^{\frac{D-1}{2}}}{\Gam\left(\frac{D-1}{2}\right)}~, 
\eeq
is the surface area of $D-2$ dimensional unit sphere. 
The $\Xi$ integral is convergent if $\RR(\nu+\kappa+D-1)<0$, which can
be satisfied since we can choose $\RR\kappa$ arbitrarily close to 
$\RR\mu$ as long as $\RR\mu<\RR\kappa(<0)$ is maintained. 
Integration over $\Xi$ leads to \eqref{formula2}.

Applying the formula \eqref{formula2} to the expression
for the generating function \eqref{generating function tochu1}, 
and replacing the integration variables $\kappa$ and $\mu$,
respectively, 
with 
\beq
w:={\kappa-\mu+\lambda\over 2}
\quad \mbox{and} \quad 
\rho:=\mu+D-1~,
\eeq
we obtain 
\beq
\mathcal{A}(\alpha_{1},\cdots,\alpha_{n})
&=&(-i)2^{2-\lambda}\pi^{\frac{D+1}{2}}\int_{w}\Gam\left[
\begin{matrix}
2w-\lambda,\, -w\\
-w+\frac{D-1}{2},\, D+\lambda,\, -\lambda,\, -\lambda,\, \lambda+1
\end{matrix}
\right]\nonumber\\
&&\quad\times 
\int_{\rho}2^{-\rho}\Gam\left[
-\lambda+\rho,\, \lambda+1+\rho,\, -\rho,\, \lambda +D-1-2w-\rho
\right]~.
\eeq

Finally, we perform the $\rho$ integration in the above expression for
$\mathcal{A}$, using the formula 
\beq\label{formula3}
\int_{\rho}2^{-\rho}\Gam[\lambda+1+\rho,\,-\lambda+\rho,\, -\rho,\, \lambda+a-1-\rho]
=\frac{2^{\lambda+a-2}}{\sqrt{\pi}}\Gam\left[-\lambda,\, \lambda+1,\, \frac{a-1}{2},\, \lambda+\frac{a}{2}\right]~,
\eeq
which can be proven as follows. 
If we close the $\rho$-path on the left hand side of \eqref{formula3} 
to the right, we have
\beq
\label{tochu3}
&&\int_{\rho}2^{-\rho}\Gam[\lambda+1+\rho,\,-\lambda+\rho,\, -\rho,\, \lambda+a-1-\rho]\nonumber\\
&&\qquad=\Gam[-\lambda,\, \lambda+1,\, \lambda+a-1]{_{2}F_{1}}\left(-\lambda,\, \lambda+1;\, 2-\lambda-a;\, \frac{1}{2}\right)\nonumber\\
&&\qquad\qquad +2^{1-\lambda-a}\Gam[2\lambda+a,\, a-1,\, 1-\lambda-a]{_{2}F_{1}}\left(2\lambda+a,\, a-1;\, \lambda+a;\, \frac{1}{2}\right)~.
\eeq
Now applying the following formulae, known respectively 
as Bailey's summation theorem and the Gauss' second summation 
theorem~\cite{Slater:1966},
\beq
&&{_{2}F_{1}}\left(\alpha,\, 1-\alpha;\, \gamma;\, \frac{1}{2}\right)=2^{1-\gamma}\sqrt{\pi}
\Gam\left[
\begin{matrix}
\gamma\\
\frac{\gamma+\alpha}{2},\, \frac{\gamma+(1-\alpha)}{2}
\end{matrix}
\right]~,\\
&&{_{2}F_{1}}\left(2\alpha,\, 2\beta;\, \alpha+\beta+\frac{1}{2};\, \frac{1}{2}\right)
=\sqrt{\pi}\Gam\left[
\begin{matrix}
\alpha+\beta+\frac{1}{2}\\
\alpha+\frac{1}{2},\,\beta+\frac{1}{2}
\end{matrix}
\right]~, 
\eeq
we obtain after simple calculations \eqref{formula3}. 
Substituting $a=D-2w$ in \eqref{formula3}, we find
\beq
\mathcal{A}(\alpha_{1},\cdots,\alpha_{n})
=(-i)(4\pi)^{D/2}\int_{w}\Gam\left[
\begin{matrix}
2w-\lambda,\, -w,\, \lambda+\frac{D}{2}-w\\
D+\lambda,\, -\lambda
\end{matrix}
\right]
F^{\lambda-2w}\left(\frac{J}{2}\right)^{2w}~.
\eeq

Recalling the definition of $F$ and $J$, \eqref{definition of C and
G}, we expand $F^{\lambda-2w}J^{2w}$ 
to be integrals with respect to the power law indices of $\alpha_{i}$'s
using Eq.~\eqref{Barnes formula2}~\cite{Marolf:2010nz}. 
Since the master integral $\mathcal{M}$ is given by the Mellin 
transform of the generating function $\mathcal{A}$, we finally obtain
\beq\label{result of the Master Integral}
\mathcal{M}(\nu_{1},\cdots,\nu_{N})
&=&
(-i)\frac{(4\pi)^{D/2}}{\Gam(D+\sum\nu_{i})\left[\prod\Gam(-\nu_{i})\right]}
\int_{(h_{ij})}\left[\prod_{i<j}\left(\frac{1-Z_{ij}}{2}\right)^{h_{ij}}\Gam(-h_{ij})\right]\nonumber\\
&&\qquad \times 
\left[\prod\Gam(H_{i}-\nu_{i})\right]\Gam\left(\frac{D}{2}+\sum\nu_{i}-\sum
h_{ij}\right)~,
\eeq
where $\int_{(h_{ij})}$ 
represents $N(N-1)/2$-hold integration 
$\prod_{1\leq i<j\leq N}\left(\int dh_{ij}/2\pi\right)$, 
and 
\beq
H_{i}:=\sum_{k=1}^{i-1}h_{ki}+\sum_{k=i+1}^{N}h_{ik}~.
\eeq

In the above derivation of the equivalence between the expressions 
\eqref{masterintegral}
and 
\eqref{result of the Master Integral} 
we assumed that all external points are mutually in spacelike
separation. 
However, we can easily extend the result 
\eqref{result of the Master Integral} 
to the case of timelike separation. 
First, notice that 
the integrand of Eq.~\eqref{masterintegral} 
is analytic for the time coordinates of the external points $\eta_{i}$ 
and also that 
this $\Omega$ integral~\eqref{masterintegral} 
continues to be well-defined and uniformly 
convergent even if $\eta_{i}$ are analytically continued to 
the region of timelike separation. 
On the other hand, the ${(h_{ij})}$ integrals in  
\eqref{result of the Master Integral} are convergent 
as long as $|\arg(1-Z_{ij})|<\pi$. 
This condition is satisfied when $\eta_{i}$ are placed on 
the original path $P$. 
Later, we need to replace $X_j$ to $Y$ and then $Y$ is placed 
on the path $P_{\svect{y}}$. 
Even in this case it can be easily verified 
that the conditions $|\arg(1-Z_{ij})|<\pi$ are satisfied.  
As a result, by the uniqueness of the analytic continuation, 
the master integral
\eqref{masterintegral}
is identical to the expression
\eqref{result of the Master Integral} 
even if the separations between some pairs of
external points are timelike. 
As a remark already mentioned at the end of the preceding subsection, 
the parameters $\nu_{1},\cdots,\nu_{n},\nu_{N}$ must satisfy the conditions
\eqref{finiteMasterIntegral} in order for 
$\mathcal{M}(\nu_{1},\cdots,\nu_{n},\nu_{N})$ to be defined.

Furthermore, without violating the convergence conditions, 
we can continue the external points in 
\eqref{result of the Master Integral} to the Euclidean region 
where $\arg(1-Z_{ij})=0$. 
Then, we find that the expression \eqref{result of the Master Integral} 
is identical 
to the one obtained in the Euclidean Field Theory in Ref.~\cite{Marolf:2010nz}, except 
for the factor of $-i$ due to convention.

\section{\label{sec5}Interacting QFT: Arbitrary Graphs}

In the preceding section, 
we computed the master integral for a massive 
scalar field using the in-in
formalism in the Lorentzian de Sitter space with the $i\epsilon$ prescription 
assuming the Euclidean vacuum at the level of non-interacting theory. 
We found that the resulting master integral is the analytic continuation
of the one computed by the Euclidean path integral. 
Then, it might be expected that these two perturbative correlators 
are equivalent to all orders of perturbation. 
We will prove this equivalence 
along our formulation in this section. 
Note that this equivalence is already shown to all orders, 
graph by graph in Ref.~\cite{Higuchi:2010xt}, in a
strictly different way from the present paper.

\subsection{Statement to be proven by induction}

It is known that the Euclidean path integral gives us a certain analytic
form corresponding to any graph
$\mathcal{V}_{N}(X_{1},\cdots,X_{N})$, which 
contributes to the $N$-pt correlator. 
The analytic expression for $\mathcal{V}_{N}$ is found in
Ref.~\cite{Marolf:2010nz} in the form
\beq
\label{Marolf and Morrison}
\mathcal{V}_{N}(X_{1},\,\cdots,\, X_{N})=\int_{(h_{ij})}\left[\prod_{i<j}^{N}\left(\frac{1-Z_{ij}}{2}\right)^{h_{ij}}\Gam(-h_{ij})\right]V_{N}(h_{ij}),
\eeq
where $V_{N}(h_{ij})$ satisfies the following properties:
\begin{itemize}
\item[1.]
The fundamental strip for each variable $h_{ij}$ of $V_{N}(h_{ij})$ contains the region
\beq\label{region of analyticity}
\RR h_{ij}\in (\sigma - \mathcal{P}_{ij}(h'),\, 0]~,
\eeq
where $\mathcal{P}_{ij}$ is a linear combination of $\RR h_{kl}$ 
excluding $\RR h_{ij}$ with non-negative coefficients
\footnote{In Ref.~\cite{Marolf:2010nz}, $\mathcal{P}_{ij}$ is set to be ``a
	 polynomial function of all $\RR h_{kl}$ except for 
	 $\RR h_{ij}$ with non-negative coefficients,'' which does not matter here.}
.
\item[2.]
When $h_{ij}$ is in the region \eqref{region of analyticity}, 
	 $V_{N}(h_{ij})$ falls off, for fixed $h_{kl}$ except for $h_{ij}$,
	 as rapidly as
\beq
V_{N}(\,\cdots,\, h_{ij}=x+iy,\,\cdots)\to e^{-\pi|y|/2}|y|^{x-1}\quad (|y|\gg 1).
\eeq
\end{itemize}
In this section we shall show 
by induction that any correlators calculated in the
in-in formalism have the same analytic form as the above obtained in the  
Euclidean path integral.

We start with some $(N+K)$-pt correlator
$\mathcal{V}_{N+K}(X_{1},\cdots,X_{N+K})$ which satisfies the properties
1 and 2 above. The succeeding steps are as follows:
\begin{itemize}
\item[(a)]Set $K$ external points, $X_{N+1},\cdots,X_{N+K}$, 
in $\mathcal{V}_{N+K}$ to $Y$.

\item[(b)]Add $M-N$ propagators connected to $Y$ and integrate over
	  $\Omega$ with respect to $Y$, 
          which gives a new $M$-point correlator with more loops.
\end{itemize}

Any graphs can be obtained by this construction, 
except for the ones containing ``one-link'' loops, 
which are to be renormalized. 
It has been already shown in Ref.~\cite{Marolf:2010nz}
that the intermediate $(N+1)$-pt function obtained in step (a) satisfies
the properties 1 and 2. Therefore, what we have to consider is step
(b). 
The resulting correlator,
$\mathcal{V}_{M}(X_{1},\cdots,X_{M})$, is given by
\beq\label{M-point}
\mathcal{V}_{M}(X_{1},\cdots,X_{M})
=\int_{\Omega}dV_{Y}\,\mathcal{V}_{N+1}(X_{1},\cdots,X_{N},Y)G(X_{N+1},Y)\cdots G(X_{M},Y).
\eeq
Integration region $\Omega$ is specified in the same manner as in Sec.~\ref{sec3.1}, 
but now with $M$ external points, $X_{1},\cdots,X_{M}$.
We show below that the $M$-pt correlator given in Eq.~\eqref{M-point}
has the form of Eq.~\eqref{Marolf and Morrison}.

\subsection{Proof}

We here set the external points in Eq.~\eqref{M-point},
$X_{1},\cdots,X_{M}$, to lie on the real Lorentzian section 
with mutually spacelike separation for technical reasons 
as in Sec.~\ref{sec4}. 
Once we succeed in proving that 
$\mathcal{V}_{M}(X_{1},\cdots,X_{M})$ 
in Eq.~\eqref{M-point} satisfies the properties 1 and 2,  
it is obvious that 
the time coordinates $\eta_{I}\,(I=1,\cdots,M)$ in 
$\mathcal{V}_{M}(X_{1},\cdots,X_{M})$ 
can be analytically continued to the timelike separation 
or the Euclidean region 
for the same reason as we discussed for $\mathcal{M}$ in 
the preceding section. 

Representing the respective factors in \eqref{M-point} 
in the Mellin-Barnes form, i.e., 
\eqref{Barnes Green's function} for $G$'s and \eqref{Marolf and
Morrison} for $\mathcal{V}_{N+1}$, we obtain
\beq
\mathcal{V}_{M}
&=&
\int_{\Omega}dV_{Y}\int_{(h_{ij})}\int_{[\nu_{i}]}\left[\prod_{i<j}^{N}\left(\frac{1-Z_{ij}}{2}\right)^{h_{ij}}\Gam(-h_{ij})\right]
\left[\prod_{i=1}^{N}\left(\frac{1-Z_{iY}}{2}\right)^{\nu_{i}}\Gam(-\nu_{i})\right]
V_{N+1}(h_{ij},\nu_{i})\nonumber\\
&&\qquad \times \int_{[\nu_{I'}]}\left[\prod_{I'=N+1}^{M}\left(\frac{1-Z_{I'Y}}{2}\right)^{\nu_{I'}}\Gam(-\nu_{I'})\psi(\nu_{I'})\right]
,
\eeq
where we have set the variables in the Barnes integral of
$G(X_{I'},Y)$'s ($I'=N+1,\cdots,M$) to $\nu_{I'}$ and those for
$\mathcal{V}_{N+1}$ to $h_{ij}\,(1\leq i< j\leq N+1)$, and we 
replaced $h_{i,N+1}\,(1\leq i\leq N)$ with $\nu_{i}$. 
Here, we have denoted in short, 
\beq
\int_{[\nu_{i}]}(\cdots):=\int_{\nu_{1}}\cdots\int_{\nu_{N}}(\cdots)~,
\eeq
and so forth. 
The integrals for
$h_{ij}$ and $\nu_{i}$ is a multiple integral and here we refer to the integration
region for them as $C$. We rewrite $\mathcal{V}_{M}$ above, using $C$, as
\beq\label{M-point tochu}
\mathcal{V}_{M}
&=&
\int_{\Omega}dV_{Y}\int_{C}\prod_{i<j}^{N}\frac{dh_{ij}}{2\pi i}\prod_{i=1}^{M}\frac{d\nu_{I}}{2\pi i}\nonumber\\
&&\qquad 
\left[\prod_{i<j}^{N}\left(\frac{1-Z_{ij}}{2}\right)^{h_{ij}}\Gam(-h_{ij})\right]
\left[\prod_{I=1}^{M}\left(\frac{1-Z_{IY}}{2}\right)^{\nu_{I}}\Gam(-\nu_{I})\right]
V_{N+1}(h_{ij},\nu_{i})\left[\prod_{I'=N+1}^{M}\psi(\nu_{I'})\right].\nonumber\\
\eeq

Now the question is whether $\Omega$ integration and $C$ integration are
exchangeable. In order to examine it, we take the absolute value
of the integrand and repeatedly integrate it to see whether the
integral is finite or not. 
Here we consider the following repeated integral
\beq
&&\int_{C}\prod\left|\frac{dh_{ij}}{2\pi i}\right|\prod\left|\frac{d\nu_{I}}{2\pi i}\right|
\left|\prod \left(\frac{1-Z_{ij}}{2}\right)^{h_{ij}}\Gam(-h_{ij})\right|
|V_{N+1}(h_{ij},\nu_{i})|\left[\prod |\psi(\nu_{I'})|\right]\nonumber\\
&&\qquad\times \left[ \int_{\Omega}|dV_{Y}|\,\left|\prod \left(\frac{1-Z_{IY}}{2}\right)^{\nu_{I}}\Gam(-\nu_{I})\right|\,\right],
\eeq
where we have dropped the indices for $\prod$ and $\sum$. 
As a default, the ranges of various indices are understood as 
$1\leq i\leq N$, $N+1\leq I'\leq M$ and $1\leq I\leq M$.

In order to evaluate $\Omega$ integration of the above expression, 
we apply the discussion in Sec.~\ref{Gen. Func. for the Master Integral}, 
but slightly modify it.
In Sec.~\ref{Gen. Func. for the Master Integral}, 
the essential point is the bound for $|\arg(1-Z_{iY})-\arg(1-Z_{NY})|$
because, in Sec.~\ref{Gen. Func. for the Master Integral}, 
the exponents, $u_{i}$, of $(1-Z_{iY})$'s 
in the integrand are not independent since $\sum^{N}u_{i}=\lambda$. 

In this subsection, however, they are mutually independent. 
Therefore, we should evaluate $|\arg(1-Z_{IY})|$ itself. 
In fact, noting that the part 
dependent on the external points of the integrand of 
the $\Omega$ integration is expressed as
\beq
\label{integrand of Omega int. for arbitrary}
\prod|(1-Z_{IY})|^{\RR \nu_{I}}\exp\left[ -\sum\arg(1-Z_{IY})\II \nu_{I} \right]
~,
\eeq
we have to bound $|\arg(1-Z_{IY})|$ for our purpose. 

Since the regions on the $\eta$-plane where $|\arg(1-Z_{IY})|=\pi$ 
are half lines going from $-\infty$ to $\eta_{I,-}$ or 
from $\eta_{I,+}$ to $+\infty$ on the real axis, 
$|\arg(1-Z_{IY})|$ is obviously less than $\pi$ if $Y\in \Omega$. 
Furthermore, since the path $P_{\svect{y}}$ on the $\eta$-plane is, 
by definition, tilted by $\epsilon$ in the far past
and deviates finitely from the region above,  
$|\arg(1-Z_{IY})|$ is bounded by $\pi-\delta'$ with $\delta'$ 
being some finite positive constant.
Therefore, we can factor out the $\II \nu_{I}$ dependent part
in Eq.~\eqref{integrand of Omega int. for arbitrary} to obtain the bound
\beq
[\mbox{Eq.}~\eqref{integrand of Omega int. for arbitrary}]
<
\prod |(1-Z_{IY})|^{\RR \nu_{I}}\exp\left[ (\pi-\delta')\sum |\II \nu_{I}| \right]~.
\eeq
Now the convergence of the volume integral follows 
in the exactly same manner as before.
Thus, we are led to discuss the following integral
\beq
\label{absolute convergence}
&&\int_{C}\prod\left|\frac{dh_{ij}}{2\pi i}\right|\prod\left|\frac{d\nu_{I}}{2\pi i}\right|
\left|\prod\left(\frac{1-Z_{ij}}{2}\right)^{h_{ij}}\Gam(-h_{ij})\right|\nonumber\\
&&\qquad \times
|V_{N+1}(h_{ij},\nu_{i})|\left[\prod |\psi(\nu_{I'})|\right]
\exp\left[(\pi-\delta')\sum |\II \nu_{I}|\right]|\Gam[-\nu_{1},\cdots,-\nu_{N}]|~.
\eeq
Recall that $\psi(\nu)$ behaves as
\beq
|\psi(x+iy)|~\to~ e^{-3\pi|y|/2}|y|^{x-1}~,\qquad (|y|\gg 1)~, 
\eeq
and 
$|\Gam(x+iy)|\approx
(2\pi)^{1/2}e^{-\pi|y|/2}|y|^{x-1/2}\,(|y|\to +\infty)$. 
Furthermore, $V_{N+1}$ behaves as
\beq
|V_{N+1}(h_{12}=x+iy,\cdots)|~\to~ e^{-\pi|y|/2}|y|^{x-1}~,
\qquad (|y|\gg 1)~, 
\eeq
from the property 2 of the assumption of induction, and 
the same is true for the other arguments, too.  
Therefore, the integral 
\eqref{absolute convergence} is convergent, and hence the order
of the integration over $\Omega$ and $C$ in \eqref{M-point tochu} are
exchangeable:
\beq\label{M-point final}
\mathcal{V}_{M}
&=&
\int_{(h_{ij})}\int_{[ \nu_{I} ]}
\left[\prod \left(\frac{1-Z_{ij}}{2}\right)^{h_{ij}}\Gam(-h_{ij})\right]
V_{N+1}(h_{ij},\nu_{i})\left[\prod \psi(\nu_{I'})\right]
\nonumber\\
&&\qquad \times\int_{\Omega}dV_{Y} 
\left[\prod_{I=1}^{M}\left(\frac{1-Z_{IY}}{2}\right)^{\nu_{I}}\Gam(-\nu_{I})\right]~.
\eeq

Now substituting the Mellin-Barnes form for the 
master integral, \eqref{result of the Master Integral}, 
into Eq.~\eqref{M-point final}, we arrive at 
the same Mellin-Barnes representation for $\mathcal{V}_{M}$ as that obtained in
Sec.~4.2 of Ref.~\cite{Marolf:2010nz}, where $\mathcal{V}_{M}$ is
shown to be represented in the form of Eq.~\eqref{Marolf and Morrison} with $V_{M}$
satisfying the properties 1 and 2. This completes the proof of the
equivalence between the two types of correlators.

\section{\label{sec6}Summary}

In this work, we considered massive interacting scalar field theory and 
demonstrated perturbative calculation for the correlators using 
in-in formalism in the flat chart of de Sitter space 
with the $i\epsilon$ prescription.
We found that the master integral defined in Eq.~\eqref{master integral} 
has completely the same Mellin-Barnes representation as that obtained in 
Ref.~\cite{Marolf:2010nz} based on the Euclidean field theory.
We then derived the analytic Mellin-Barnes 
formulae for the correlators of quantum field on the flat chart. 
The resulting correlators are shown to be 
completely the same as the analytic continuations of 
the ones considered in the Euclidean field theory. 
Thus we find that the $i\epsilon$ prescription 
in de Sitter space gives the Euclidean vacuum.

Although the relation between these two vacua has been clarified in 
Ref.~\cite{Higuchi:2010xt}, in order to extend this to massless field theory,
we gave an alternative proof of their equivalence by direct calculation.
In particular, graviton in de Sitter space has been 
a topic of much discussion. (See, e.g. \cite{Tsamis:1994ca, Tsamis:1996qq, 
Garriga:2007zk, Higuchi:2011vw}.) 
It is also worth considering
derivatively interacting massless scalar field as a model of graviton.

The proof in Ref.~\cite{Higuchi:2010xt} of the equivalence between the two vacua 
relies on the decay of the propagator at a large separation. But the propagators 
in massless theory do not fall off in general. 
This could be an obstacle in extending the discussion 
to interacting massless field theory.
Though we considered only massive theory in this work,
we believe that our proof has potential to be extended to wider range of theories
which include derivatively interacting massless field theory, 
since our proof of the correspondence of the correlators is based on
direct calculation without relying on this property.

\section*{Acknowledgments}

We would like to thank H.~Kitamoto 
and V.~Onemli for valuable comments. 
YK also thanks to R.~Saito, K.~Sugimura 
and K.~Nakata for useful and interesting discussions. 
YK is supported by the Grant-in-Aid for JSPS Fellows No.~24-4198.
TT is supported by the Grand-in-Aid for Scientific Research 
Nos.~21111006, 21244033, 24103001 and 24103006.
This work was also supported by the Grant-in-Aid for the Global COE
programs, ``The Next Generation of Physics, Spun from Universality
and Emergence'' from the Ministry of Education, Culture, Sports, 
Science and Technology of Japan.

\appendix

\section{Formula}

Let $A_{1},\cdots,A_{n+1}$ be complex numbers satisfying $|\arg
A_{i}-\arg A_{j}\,|<\pi\,\,(\forall\,i,j)$. Then, the following 
formula is true as a repeated integral and also as a
multiple integral 
since the integral is easily shown to be
independent of the order of the integration: 
\beq\label{Barnes formula2}
&&(A_{1}+A_{2}+\cdots +A_{n+1})^{\lambda}\nonumber\\
&&\qquad =\frac{1}{\Gam(-\lambda)}\int_{u_{1}}\cdots \int_{u_{n}}\Gam\left[-\lambda +\sum u_{i},\, -u_{1},\,\cdots ,\, u_{n}\right]\left(A_{1}\right)^{u_{1}}\cdots \left(A_{n}\right)^{u_{n}}\left(A_{n+1}\right)^{\lambda-\sum u_{i}}.
\eeq

\noindent
Proof of \eqref{Barnes formula2}: 

The basic formula is the following:
\beq\label{Barnes formula1}
(a+b)^{\lambda}=\frac{1}{\Gam(-\lambda)}\int_{\mu}\Gam[-\lambda +\mu,\, -\mu]a^{\mu}b^{\lambda -\mu}~,
\quad (|\arg a-\arg b\,|<\pi)~.
\eeq
One applies this formula \eqref{Barnes formula1} with $a=A_{n},\,
b=A_{1}+\cdots+A_{n-1}+A_{n+1}$, and then again apply 
\eqref{Barnes formula1} to $(A_{1}+\cdots+A_{n-1}+A_{n+1})^{\lambda-\mu}$
in the result of the previous step with 
$a=A_{n-1},\, b=A_{1}+\cdots+A_{n-2}+A_{n+1}$. Repeating the same 
operation, one formally reaches \eqref{Barnes formula2}. The point is
that the conditions 
\beq
&&|\arg A_{1}- \arg A_{n+1}\,|<\pi~,\nonumber\\
&&|\arg A_{2}-\arg(A_{1}+A_{n+1})\,|<\pi~,\nonumber\\
&&\qquad\cdots\nonumber\\
&&|\arg A_{n}-\arg(A_{1}+\cdots+ A_{n-1}+A_{n+1})\,|<\pi~, 
\eeq
are required to perform the above transformation. 
To allow the exchange of the order of the repeated
integration without changing the result, we impose 
stronger conditions
\beq\label{sufficient condition for commutation}
&&|\arg A_{P(1)}- \arg A_{n+1}\,|<\pi~,\nonumber\\
&&|\arg A_{P(2)}-\arg(A_{P(1)}+A_{n+1})\,|<\pi~,\nonumber\\
&&\qquad\cdots\nonumber\\
&&|\arg A_{P(n)}-\arg(A_{P(1)}+\cdots+ A_{P(n-1)}+A_{n+1})\,|<\pi~,
\eeq
for any permutation $P$. 
It is easily verified that, if we choose $A_{i}$'s to satisfy 
\begin{equation}
 |\arg A_{i}-\arg A_{j}\,|=|\arg(A_{i}/A_{j})|<\pi/2~,
\end{equation}
for all pairs of $i$ and $j$, the conditions 
\eqref{sufficient condition for commutation} are all satisfied 
because in general $|\arg(\sum_I r_I)|<\pi/2$ if $|\arg(r_I)|<\pi/2$ 
for all $r_I$ where $r_{I}$'s are understood as $A_{i}/A_{j}$'s. 
However, this restriction can be easily relaxed by 
analytic continuation with respect to $A_i$ as long as 
the conditions  
\begin{equation}
 |\arg A_{i}-\arg A_{j}|<\pi~,
\end{equation}
are satisfied for all pairs since the right hand side of 
Eq.~\eqref{Barnes formula2} continues to converge 
under these conditions. 
This completes the proof of \eqref{Barnes formula2}.

\bibliographystyle{myJHEP}
\bibliography{deSitter.bib}

\end{document}